\documentclass[onecolumn]{quantumarticle}

\usepackage[utf8]{inputenc}
\usepackage[english]{babel}
\usepackage[T1]{fontenc}
\usepackage{amsmath}
\usepackage{hyperref}
\usepackage{tikz}
\usepackage{lipsum}

\usepackage{mathtools}
\usepackage{amsfonts}
\usepackage{amsmath}
\usepackage{graphicx}
\usepackage{amssymb}
\usepackage{longtable}
\usepackage{cancel}
\usepackage{bm} 
\usepackage{bbm} 
\usepackage{empheq} 

\newcommand{\beq}{\begin{equation}} 
\newcommand{\eeq}{\end{equation}}
\newcommand{\bqa}{\begin{eqnarray}} 
\newcommand{\eqa}{\end{eqnarray}}
\newcommand{\nn}{\nonumber}

\newcommand{\dg}{^\dagger}
\newcommand{\rt}[1]{\sqrt{#1}\,}

\newcommand{\bra}[1]{\langle{#1}|} 
\newcommand{\ket}[1]{|{#1}\rangle}

\newcommand{\op}[2]{\left|{#1}\rangle \langle{#2}\right| }

\newcommand{\tbt}[4]{\left(\begin{array}{cc} {#1}& {#2} \\ {#3}&{#4} \end{array}\right)}

\newcommand{\sch}{Schr\"odinger}

\newcommand{\Qhat}{\hat{Q}}

\newcommand{\dt}{\delta t}
\newcommand{\Ls}{{\cal L}}
\newcommand{\Ks}{{\cal K}}
\newcommand{\Ds}{{\cal D}}

\newcommand{\rhoi}{\rho_{\rm i}}
\newcommand{\rhof}{\rho_{\rm f}}
\newcommand{\Lstar}{\bar{\Ls}_\star}
\newcommand{\Jstar}{{\cal J}_\star}
\newcommand{\psiN}{\ket{\psi_N}}
\newcommand{\psiNp}{\ket{\psi_{N+1}}}

\begin{document}

\title{Hitting statistics from quantum jumps}
\date{\today}

\author{A. Chia}
\affiliation{Centre for Quantum Technologies, National University of Singapore}

\author{T. Paterek}
\affiliation{Division of Physics and Applied Physics, School of Physical and Mathematical Sciences, Nanyang Technological University, Singapore}
\affiliation{Majulab, CNRS-UNS-NUS-NTU International Joint Research Unit, UMI 3654, Singapore}

\author{L. C. Kwek}
\affiliation{Centre for Quantum Technologies, National University of Singapore}
\affiliation{Majulab, CNRS-UNS-NUS-NTU International Joint Research Unit, UMI 3654, Singapore}
\affiliation{Institute of Advanced Studies, Nanyang Technological University, Singapore}
\affiliation{National Institute of Education, Nanyang Technological University, Singapore}

\maketitle

\begin{abstract}

We define the hitting time for a model of continuous-time open quantum walks in terms of quantum jumps. Our starting point is a master equation in Lindblad form, which can be taken as the quantum analogue of the rate equation for a classical continuous-time Markov chain. The quantum jump method is well known in the quantum optics community and has also been applied to simulate open quantum walks in discrete time. This method however, is well-suited to continuous-time problems. It is shown here that a continuous-time hitting problem is amenable to analysis via quantum jumps: The hitting time can be defined as the time of the first jump. Using this fact, we derive the distribution of hitting times and explicit exressions for its statistical moments. Simple examples are considered to illustrate the final results. We then show that the hitting statistics obtained via quantum jumps is consistent with a previous definition for a measured walk in discrete time [Phys.~Rev.~A~{\bf 73}, 032341 (2006)] (when generalised to allow for non-unitary evolution and in the limit of small time steps). A caveat of the quantum-jump approach is that it relies on the final state (the state which we want to hit) to share only incoherent edges with other vertices in the graph. We propose a simple remedy to restore the applicability of quantum jumps when this is not the case and show that the hitting-time statistics will again converge to that obtained from the measured discrete walk in appropriate limits.

\end{abstract}

\section{Introduction}
\label{Intro}

With the advent of quantum information science and the desire to build a quantum computer, the study of quantum algorithms have become an integral part of quantum information theory \cite{NC00}. Quantum walks have played a special role in quantum computing by providing a platform on which quantum algorithms may be analysed \cite{VA08,Por13}. Moreover, they have served as a useful mechanism to describe and explain coherent transport processes in photosynthesis \cite{ECR+07,MRLAG08} and the breakdown of a driven system in an electric field \cite{OKAA05}. This has in turn stimulated much experimental effort to realise quantum walks, see e.g. Refs.~\cite{PLP+08,KFC+09,SMS+10,ZKG+10}. Theoretical quantum optics on the other hand has had fruitful applications in the analyses of quantum technologies \cite{BEZ01,KL10,WM10}. In this paper we will apply the well-known theory of quantum jumps developed in quantum optics to define and calculate the distribution of hitting times in open (i.e.~non-unitary) quantum walks.

The paper is organised as follows. We first review quantum jumps and motivate its application to hitting problems in quantum walks in Sec.~\ref{QuantumJumps}. The theory of quantum walks and some existing works related to quantum jumps are then reviewed in Sec.~\ref{QuantumWalks}. We then introduce the necessary background in Sec.~\ref{Background} with the presentation of the quantum-jump formalism in Sec.~\ref{MCW}, followed by our approach to continuous-time open quantum walks in Sec.~\ref{QWDefined}. In Sec.~\ref{HTDdefn} we explain how the quantum-jump method is to be applied to quantum walks and obtain our first result---the hitting-time distribution. Then in Sec.~\ref{AvgHittingTime} we derive explicit expressions for the hitting-time statistics. We then relate our result to a previous definition of the hitting time devised for a discrete-time measured walk in Sec.~\ref{SEC_COMPARISON}. Here the hitting-time statistics of the discrete-time measured walk is shown to converge to the quantum-jump approach in the continuous-time limit. A problem with the quantum-jump definition of hitting times is that it becomes inaccurate when there are coherent transitions to the final state. We overcome this problem in Sec.~\ref{N+1Model}, extending the quantum-jump approach to arbitrary graphs. We then conclude with a summary of our results and analyses in Sec.~\ref{Conclusion} and also comment on the relationship of our work with other studies not mentioned in the literature review of Secs.~\ref{QuantumJumps} and \ref{QuantumWalks}.

\subsection{Quantum jumps---from photon counting to hitting times}
\label{QuantumJumps}

\begin{figure}[t]
\centerline{\includegraphics[width=14cm]{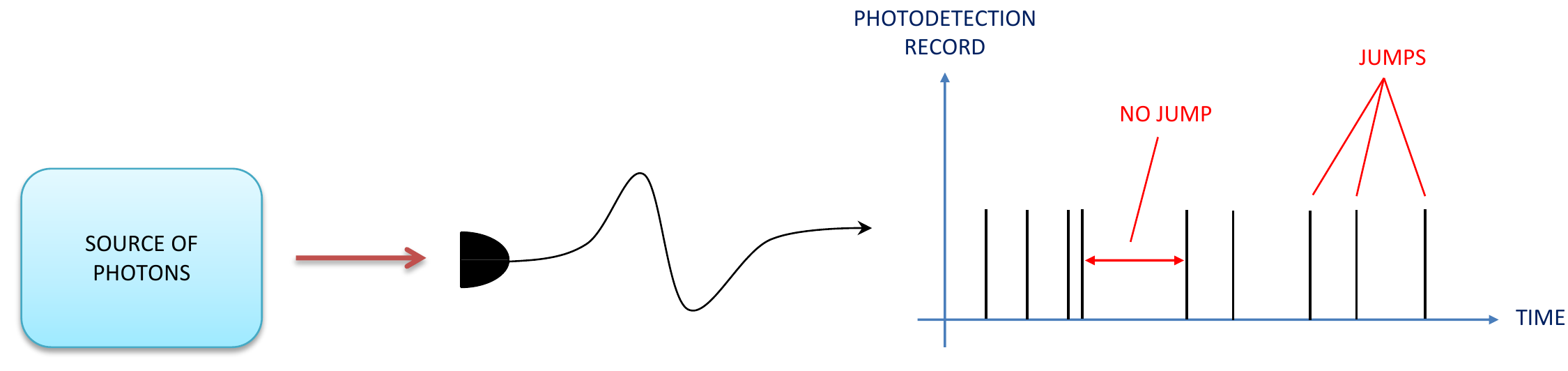}}
\caption{\label{Photodetection} Typical scenario where quantum-jumps are applied in quantum optics. Usually an atomic system or an optical cavity with a lossy end mirror acts as a source of photons. The output light is measured by a photodetector. It is well known in quantum optics that the photodetection record contains information about the state of the light. The photodetection record can also be used to gain information about the photoemissive source. References \cite{HYK97,GW01,DFVKW08,KM15} are some examples.}
\end{figure}

The quantum-jump approach to dissipative quantum dynamics has traditionally been used for efficiently solving master equations \cite{DCM92,DZR92,DPZG92,MCD93}, or calculating photon statistics in photon counting \cite{CSVR89,Car93}. It also goes by the name of Monte Carlo wavefunctions or quantum trajectories due to the different contexts in which it was invented (although quantum jumps correspond only to a subset of unravellings within quantum-trajectory theory\footnote{We have devoted a separate discussion to the use of other unravellings in hitting problems in Sec.~\ref{FurtherDiscussions}. If the reader is already familiar with quantum-trajectory theory then this may be read now, but otherwise should be left till the end for the nonexperts.}). We refer the reader to Ref.~\cite{PK98} for a comprehensive review of how the quantum-jump method was developed. Figure \ref{Photodetection} illustrates a typical scenario where quantum jumps are applied e.g.~Refs.~\cite{HYK97,GW01,DFVKW08,KM15}. There is usually a system (described by a master equation \cite{Car08}), say a lossy cavity or a two-level atom which dissipates energy into the environment in the form of photons. The emitted photons are then measured by a photodetector. Each count, or photodetector ``click'' appears as a spike in the photodetection record. The quantum-jump formalism then allows one to compute the system evolution conditioned on a given photodetection record where each detector click is associated with a ``jump'' in the system Hilbert space. Periods of no clicks then correspond to system evolution with no jumps. The quantum-jump method states that an average over a large number of such conditioned states will reproduce the solution to the master equation. Thus in solving the master equation, one also gains access to the statistics of photon arrival times at the detector. Photon statistics are therefore often calculated using the quantum-jump method \cite{KM15,JGCRP99,KLM13}. In fact, Carmichael derived the quantum-jump method \cite{CSVR89,Car93} from the photon-counting distribution due to Kelley and Kleiner \cite{KK64}. It is this intuition between photodetector clicks and quantum jumps that we will  exploit for deriving the hitting-time statistics in quantum walks. To see this we need to first explain what hitting times are: In quantum-walk theory the system, usually called the walker, has a Hilbert space spanned by a countable set of orthonormal states $\{\ket{\psi_n}\}_n$, called sites, nodes, or vertices. The hitting time is defined as the time required for the walker to reach (or ``hit'') some ``final'' state $\ket{\psi_N}$ for the first time given that initially it was at $\ket{\psi_m}$. The hitting time is a random variable since the first time that the walker arrives at $\ket{\psi_N}$ will vary from one realisation of the quantum walk to another. It therefore makes sense to speak of the average hitting time and its higher-order statistics. We will often call the hitting-time distribution simply as the hitting distribution for ease of reference. Similarly, hitting statistics (as in the title) refers to the statistics of hitting times. It should be noted that the problem we have defined here has an analogue in classical Markov chains, where the orthonormal set $\{\ket{\psi_n}\}_n$ is replaced by a set of probabilities. The temporal evolution of the classical walker is then governed by a set of rate equations for the site probabilities \cite{KMT12} which plays the role of the master equation in the quantum case. In fact, one of the quantum-walk models that we will discuss two paragraphs down is based on this idea. We should also mention that hitting times are usually called first-passage times in the classical theory of Markov chains and most areas of science where this concept appears, see e.g.~Refs.~\cite{Red01,TM12,CBTVK07}. 

\begin{figure}[t]
\centerline{\includegraphics[width=14cm]{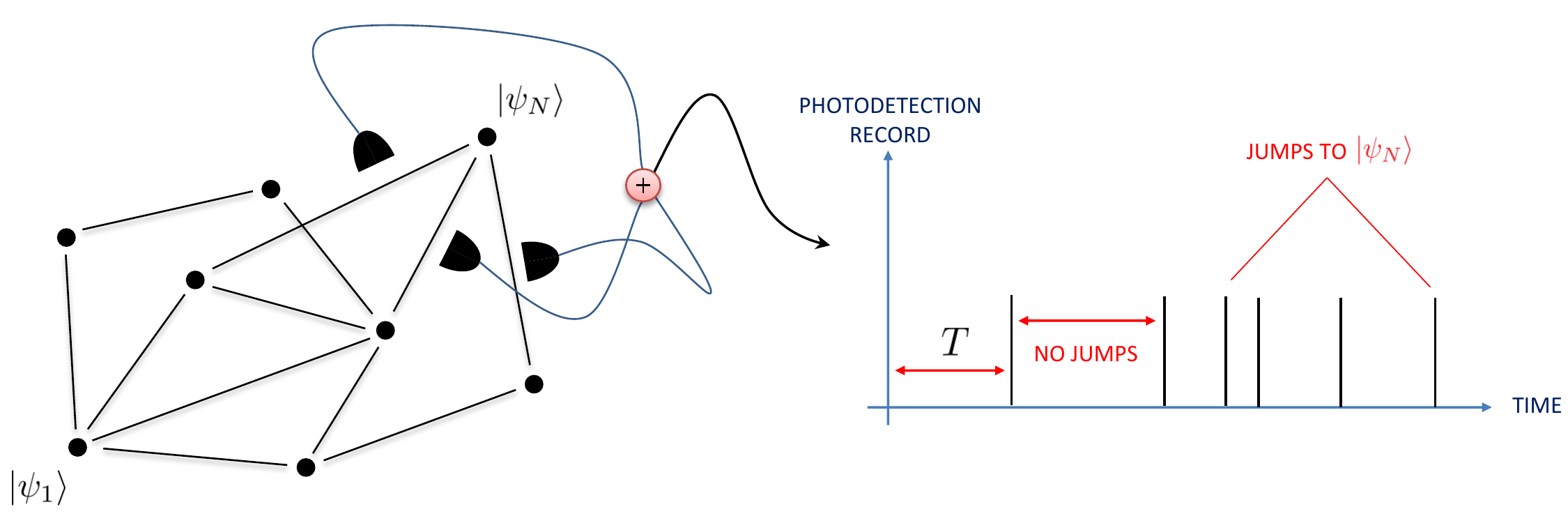}}
\caption{\label{JumpsOnGraph} An example graph to illustrate the quantum-jump approach to hitting times. We place imaginary detectors that can only detect transitions to $\ket{\psi_N}$ along the edges connected to $\ket{\psi_N}$. We then lump all the detection records obtained from these photodetectors into one record. A click in the superimposed record therefore corresponds to a transition to $\ket{\psi_N}$. This can be described by a jump in the Hilbert space of the walker as prescribed by the quantum-jump method. When viewed as such, the hitting time, labelled $T$, is therefore the time at which the first click occurs.}
\end{figure}

The quantum-jump method lends itself naturally to a hitting problem because the question one is asking, starting from the initial time, is whether the system has made a transition to the final state or not. This leads to a very natural division of the system evolution into two types---either a transition to the final state occurred, or it has not, and this corresponds to a system evolution conditioned on ``jumps'' and ``no jumps''. We illustrate this idea on an arbitrarily chosen quantum walk in Fig.~\ref{JumpsOnGraph}. The walker can move between any two vertices connected by a line, usually called an edge (edges and vertices together form a graph as shown). We are not defining the edges precisely for the moment but one may think of each edge as given by a completely-positive trace-preserving map, i.e.~a Kraus channel of some sort \cite{NC00,Kra71}. Consider the hitting problem defined by the initial and final states shown as $\ket{\psi_1}$ and $\ket{\psi_N}$ respectively on the graph. Imagine now that each transition on the graph emits a photon. Then we can find out when the quantum walker will arrive at $\ket{\psi_N}$ by associating photodetectors with the edges connected to $\ket{\psi_N}$ (assuming our photodetectors only measure photons emitted from transitions to $\ket{\psi_N}$). If we superimpose the photodetection records from these detectors then we will observe an overall record shown on the right of Fig.~\ref{JumpsOnGraph}. In this case a photodetector click signifies that our walker is at $\ket{\psi_N}$, and the first such click provides the hitting time. We will not be able to distinguish which channel the quantum walker took to reach $\ket{\psi_N}$ but this does not matter if we only care about when the walker visits $\ket{\psi_N}$. This analogy makes clear how the quantum-jump method can be applied to calculate the distribution of hitting times in quantum walks. The theory of quantum walks is more than two decades old so let us briefly review some of the key developments in this field and then discuss where our work stands in relation to the literature on this topic. 

\subsection{Quantum walks and quantum trajectories}
\label{QuantumWalks}

The concept of a quantum walk was first introduced by Aharonov and coworkers in 1993 \cite{ADZ93}. Quantum walks can be categorised into one of two classes---either a walk in discrete time, or a walk in continuous time \cite{VA12}. Quantum walks in discrete time follow the idea originally prosposed by Aharonov and colleagues where a quantum coin is introduced. Classically, a random walk on a line involves a coin toss, the result of which determines whether the walker moves one step to the left, or one step to the right. In the quantum version the coin is simply a two-state system. The movement of the walker is then effected by a (quantum) coin toss implemented as a Hadamard gate $\hat{C}$, followed by applying a shift operator $\hat{S}$ to change the walker's position conditioned on the coin state \cite{Kem03a}. Each application of the unitary operator $\hat{U}\equiv\hat{S}\hat{C}$ then evolves the walker in one time step. For a unitary quantum walk, the notion of hitting becomes fuzzy due to the walker's ability to be in a superposition of sites. As a result, multiple definitions of hitting have been proposed \cite{Kem03b,KB06}. Among these, the definition of hitting based on repeated measurements proposed by Krovi and Brun in Ref.~\cite{KB06} will be useful for our analysis. Applications of discrete-time quantum walks include the design of search algorithms \cite{SKW03,AKR05}, and implementing universal quantum computation \cite{LCETK10}. Continuous-time quantum walks on the other hand do not use a coin. Instead, the quantum walker evolves according to a unitary operator generated by a Hamiltonian, i.e.~$\hat{U}(t)=\exp(-i\hat{H}t)$ with $\hat{H}$ parametrised by hopping rates to adjacent nodes. Quantum walks in continuous time were first introduced by Farhi and Gutmann \cite{FG98} in close analogy to continuous-time Markov chains \cite{FG98,CFG02}. Their hitting properties were then studied by Varbanov and colleagues \cite{VKB08} who defined hitting times through jumplike weak measurements. The quantum-jump treatment in the present paper can also be seen as a weak measurement of the walker position in that most of the time one gets a null result (no jumps) which does not modify the walker state very much but once in a while a collapse (a jump) happens which changes the walker state drastically \cite{Bru02}. As with the discrete-time case, continuous-time quantum walks have also been shown to give rise to exponential speedups over classical algorithms for searching \cite{CG04} and solving black-box problems \cite{CCD+03}. It has also been shown that universal quantum computation can be achieved using continuous-time quantum walks \cite{Chi09}. Discrete-time quantum walks were then generalised to allow for non-unitary evolution. These are known as open quantum walks and were first introduced by Attal and collaborators \cite{APS12,APSS12}. Here the coin changes its state according to a completely-positive trace-preserving map and the evolution operator $\hat{U}$ for the unitary walk is also replaced by a completely-positive trace-preserving map. A continuous-time model was subsequently proposed by Pellegrini as the continuous-time limit of the discrete-time open quantum walk \cite{Pel14}. Most recently, Liu and Balu proposed yet another continuous-time model of open quantum walks by starting with a master equation in the Lindblad form \cite{LB16}. This approach to defining quantum walks is in the same spirit as Farhi and Gutmann's for unitary walks as already alluded to above. They (Farhi and Gutmann) viewed the \sch\ equation as the analogue of the rate equation for probabilities in classical Markov chains. Generalising this to allow for non-unitary dynamics naturally leads to a master equation. We also consider here continuous-time open quantum walks described by a master equation.

\begin{figure}[t]
\centerline{\includegraphics[width=14cm]{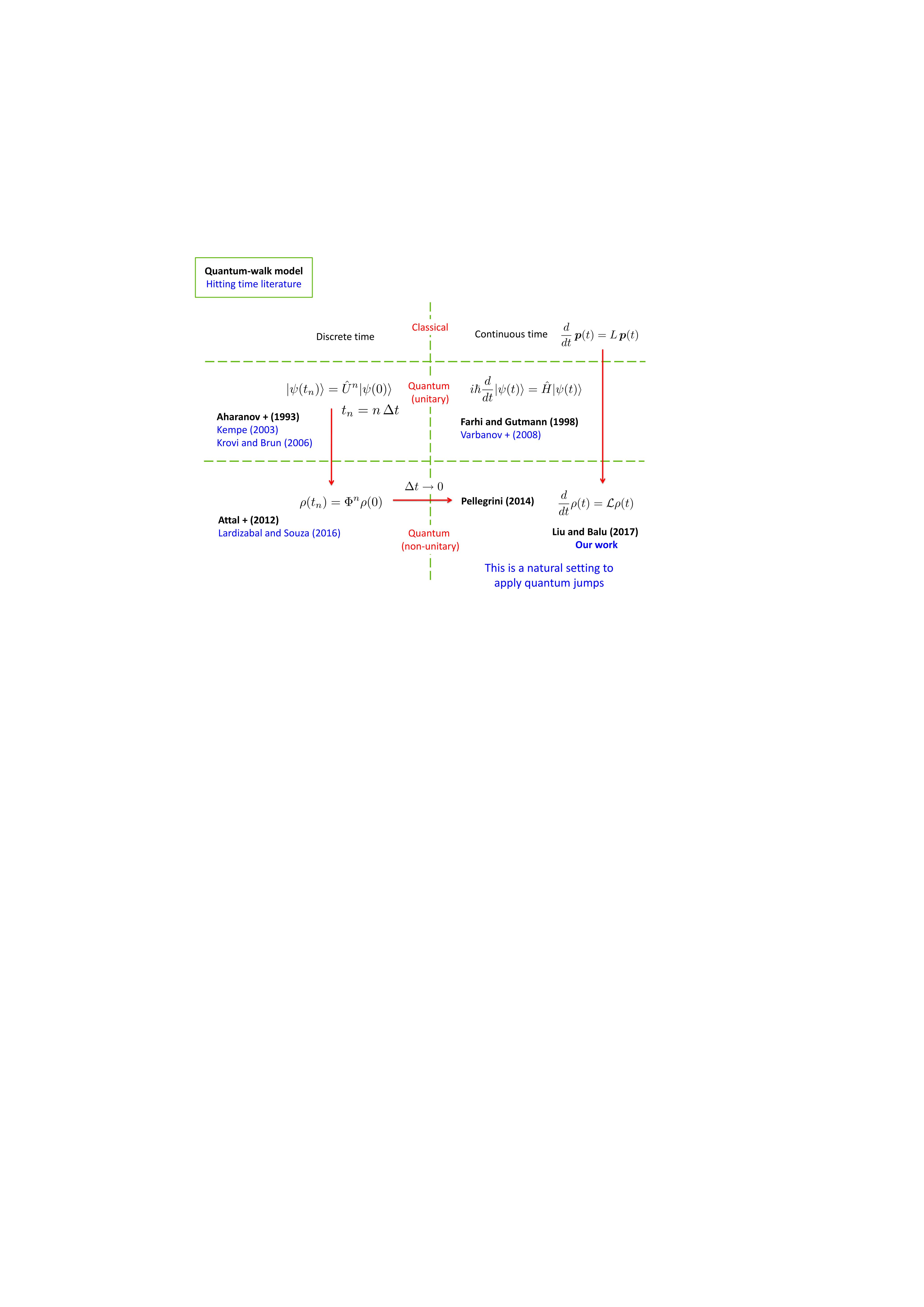}}
\caption{\label{Literature} Representative works of different quantum-walk models and their corresponding hitting-time proposals: Aharanov\,+ (1993) Ref.~\cite{ADZ93}; Kempe (2003) Ref.~\cite{Kem03b}; Krovi and Brun (2006) Ref.~\cite{KB06}; Farhi and Gutmann (1998) Ref.~\cite{FG98}; Varbanov\,+ (2008) Ref.~\cite{VKB08}; Attal\,+ (2012) Refs.~\cite{APS12,APSS12}; Lardizabal and Souza (2017) Ref.~\cite{LS16}; Liu and Balu (2017) Ref.~\cite{LB16}; Pelligrini (2014) Ref.~\cite{Pel14}.}
\end{figure}

Of particular interest for our work is the application of quantum trajectories in Refs.~\cite{APS12,APSS12,Pel14}. These works were motivated by the fact that quantum trajectories offer an efficient means of simulating quantum walks and have the advantage of providing a visualisation of the walk. Lardizabal and Souza have recently defined hitting times using quantum trajectories \cite{LS16}, but for the discrete-time model due to Attal and colleagues \cite{APS12,APSS12}. Given that quantum trajectories were invented in the context of Lindblad-form master equations the latter is in fact the most natural setting in which to consider them. For such non-unitary graphs one can therefore bypass the formalism of positive operator-valued measures (as in Ref.~\cite{VKB08} for example) and give a quantum-trajectory treatment of the hitting problem. Our aim in this paper is to define the hitting time for open quantum walks as the time for the first jump to occur and use this as the basis for calculating the distribution of hitting times and its statistical moments. Our result also corrects Ref.~\cite{QDX12} which claims to have calculated the average hitting time for continuous-time open quantum walks. However, an inspection of Ref.~\cite{QDX12} shows their model to be purely classical. A summary of the relevant literature is given in Fig.~\ref{Literature}.

\section{Background}
\label{Background}

\subsection{Quantum jumps}
\label{MCW}

Here we will explain the method of quantum jumps as illustrated in Fig.~\ref{Photodetection}, i.e.~in the context of photon counting. This makes the basic structure of the formalism intuitive. Assume for simplicity that the system in Fig.~\ref{Photodetection} is a single-mode field in an optical cavity with one slightly lossy mirror. The lossy mirror allows the single-mode light to couple to the field outside the cavity which we assume is in a vacuum state. This is described by a master equation which has only one dissipative channel, given by
\begin{align}
\label{L}
	\dot{\rho}(t) = {\cal L} \rho(t) \equiv -i \big[\hat{H},\rho(t)\big] + \hat{c} \, \rho(t) \, \hat{c}\dg - \frac{1}{2} \, \big\{ \hat{c}\dg \hat{c}, \rho(t) \big\} \;,
\end{align}
where $\hat{c}=\rt{\gamma}\hat{a}$ with $\gamma$ being a damping coefficient and $\hat{a}$ a bosonic annihilation operator. We are using the notation $\{\hat{A},\hat{B}\}$ for the anticommutator of any two operators $\hat{A}$ and $\hat{B}$. The terms containing $\hat{c}$ modify the usual unitary dynamics (which conserves the system's energy) to include a dissipative process (here being the loss of photons to the vacuum outside the optical cavity). The quantum-jump formalism provides a way to understand these terms as follows. The system state $\rho(t)$ can evolve over an infinitesimal time interval $dt$ conditioned on two types of detection events---a count, and no count (Fig.~\ref{Photodetection}). Given an initial state $\rho(t)$, the probability of registering a count in the interval $[t,t+dt)$ is given by
\begin{align}
\label{ProbJump}
	\wp_1(dt) = {\rm Tr}\big\{ {\cal J} \rho(t) \big\} dt = {\rm Tr}\big\{ \hat{c}\dg \hat{c} \, \rho(t) \big\} dt  \;,   
\end{align}
where we have defined
\begin{align}
	{\cal J} \rho \equiv \hat{c} \, \rho \, \hat{c}\dg  \;.
\end{align}
Here we are assuming our detector in Fig.~\ref{Photodetection} is ideal. The state change conditioned on observing a count (a spike in the photodetection record in Fig.~\ref{Photodetection}) is effected by
\begin{align}
\label{Jump}
	\rho_1(t+dt) = \frac{{\cal J} \rho(t)}{{\rm Tr}\big\{ {\cal J} \rho(t) \big\}}   \;. 
\end{align}
Equation \eqref{Jump} is called a quantum jump and $\hat{c}$ is referred to as a jump operator. A quantum jump causes $\rho(t)$ to change discontinuously in time.

The second type of evolution is conditioned on the event that no counts were registered in $[t,t+dt)$. This is commonly referred to as the no-jump evolution and occurs with probability
\begin{align}
\label{ProbNoJump}
	\wp_0(dt) = 1 - \wp_1(dt)  \;.
\end{align}
The state at time $t+dt$ conditioned on a no-count observation is
\begin{align}
\label{NoJump}
	\rho_0(t+dt) = \frac{e^{\bar{\cal L}dt} \rho(t)}{{\rm Tr}\big[ e^{\bar{\cal L}dt} \rho(t) \big]}  \;,
\end{align}
where 
\begin{align}
\label{NoJumpGenerator}
	\bar{\cal L} \, \rho(t) \equiv  -i \big[\hat{H},\rho(t)\big] - \frac{1}{2} \, \big\{ \hat{c}\dg \hat{c},\rho(t) \big\}  \;.
\end{align}
Note that $\wp_0(dt)$ can be written explicitly in terms of \eqref{NoJumpGenerator} as
\begin{align}
\label{wp0}
	\wp_0(dt) = {\rm Tr}\big[ e^{\bar{\cal L} dt} \rho(t) \big]  \;.
\end{align}
It helps to consider two generalisations of the basic structure given by \eqref{ProbJump}--\eqref{NoJumpGenerator} for the application of quantum jumps to quantum walks coming up later.

\subsubsection{Generalisation 1: Non-unit detection efficiency}

The above formulation assumes that every jump in the system is observed with probability one. Therefore the first generalisation that we will consider is to allow for an non-ideal observation, or non-unit detection efficiency where only a fraction $\eta$ (between zero and one) of the jumps are actually detected \cite{WM10}. This means that the probablity of detecting a jump in time $dt$ (when it occurs) should be \eqref{ProbJump} multiplied by $\eta$. This can be effected in the quantum-jump approach by letting 
\begin{align}
	{\cal J} \rho(t) \; \longrightarrow \; {\cal J}(\eta) \;\!\rho(t)  \equiv \eta \, \hat{c} \, \rho(t) \, \hat{c}\dg  \;.
\end{align}
The probability of detecting a jump in the time interval $[t,t+dt)$ is now
\begin{align}
	\wp_1(\eta;dt) = \eta \, {\rm Tr}\big[ \hat{c}\dg \hat{c} \, \rho(t) \big] dt  \;.
\end{align}
The remaining fraction of jumps that went unnoticed then contributes to the no-count evolution and we have
\begin{align}
	\bar{\cal L} \rho(t) \; \longrightarrow \; 
	\bar{\cal L}(\eta) \, \rho(t) \equiv -i \big[\hat{H},\rho(t)\big] - \frac{1}{2} \, \big\{ \hat{c}\dg \hat{c}, \rho(t) \big\} + (1-\eta) \, \hat{c} \, \rho(t) \, \hat{c}\dg  \;.
\end{align}
It is not difficult to show that this gives the probability of no counts in $[t,t+dt)$ to be
\begin{align}
	\wp_0(\eta;dt) = {\rm Tr}\big[ e^{\bar{\cal L}(\eta) dt} \rho(t) \big] = 1 - \wp_1(\eta;dt)  \;,
\end{align}
as one would expect.

\subsubsection{Generalisation 2: Multiple decay channels}
\label{MultichannelQuantumJumps}

The second generalisation that we shall consider is to include the possibilility of multiple decay channels in the system. Assuming there are $L$ such channels, the master equation in \eqref{L} is generalised to 
\begin{align}
\label{MultichannelL}
	\dot{\rho}(t) = -i \big[\hat{H},\rho(t)\big] 
	                + \sum_{k=1}^L  \bigg[ \hat{c}_k \, \rho(t) \, \hat{c}\dg_k - \frac{1}{2} \, \big\{ \hat{c}\dg_k \hat{c}_k, \rho(t) \big\} \bigg]  \;.
\end{align}
If we now assume that each channel has a detection efficiency $\eta_k$ ($k=1,2,\ldots L$), the new superoperator effecting a jump conditioned on a detection in $[t,t+dt)$ is given by
\begin{align}
\label{JumpEta}
	{\cal J}({\bm \eta}) \, \rho(t) \equiv \sum_{k=1}^L  \, \eta_k \, \hat{c}_k \, \rho(t) \, \hat{c}\dg_k  \;, 
\end{align}
where we have defined ${\bm \eta}\equiv (\eta_1,\eta_2,\ldots,\eta_L)$. The probability of observing a jump in any one of the $L$ channels in duration $dt$ is 
\begin{align}
\label{MultichannelCountProb}
	\wp_1({\bm \eta};dt) = \sum_{k=1}^L \, \eta_k \, {\rm Tr}\big[ \hat{c}\dg_k \hat{c}_k \, \rho(t) \big] \, dt  \;.
\end{align}
The no-count superoperator is now
\begin{align}
\label{MultichannelNoCountL}
	\bar{\cal L}({\bm \eta}) \, \rho(t) 
	\equiv -i \big[\hat{H},\rho(t)\big] + \sum_{k=1}^L \bigg[ (1-\eta_k) \, \hat{c}_k \, \rho(t) \, \hat{c}\dg_k - \frac{1}{2} \, \big\{ \hat{c}\dg_k \hat{c}_k, \rho(t) \big\} \bigg] \;,
\end{align}
and the corresponding probability for no detections in the interval $[t,t+dt)$ is 
\begin{align}
	\wp_0({\bm \eta};dt) = {\rm Tr}\big[ e^{\bar{\cal L}(\bm \eta) dt} \rho(t) \big] = 1 - \wp_1({\bm \eta};dt)  \;.
\end{align}
The generalisation to multiple decay channels allows us to selectively tune the detection efficiency of channel $k$ by choosing a value for $\eta_k$. In particular we can choose to ignore jumps for selected channels by setting the detection efficiency for those channels to zero. The multichannel theory of quantum jumps with non-unit detection efficiency has been used to show that quantum jumps are more robust to measurement inefficiencies in disproving the existence of objective pure-state dynamical models \cite{DW14}. The same formalism has also been used to study parameter estimation from multichannel photon counting\footnote{Analogies between photon counting and quantum walks in connection to Ref.~\cite{KM15} are discussed further in Sec.~\ref{FurtherDiscussions}.} \cite{KM15}.

\subsection{Open quantum walks}
\label{QWDefined}

We now define the problem to which we would like to apply the quantum-jump formalism. Our problem can be set up in analogous fashion to a classical Markov chain problem in continuous time. We are given a quantum system that makes transitions between a set of discrete states
\begin{align}
\label{SpaceS}
	\mathbb{S}_N \equiv \{ \ket{\psi_1},\ket{\psi_2},\ldots,\ket{\psi_N} \}  \;.
\end{align}
The set $\mathbb{S}_N$ forms an orthonormal basis for the system. The space in which the system resides is thus spanned by $\mathbb{S}_N$. In the language of quantum-walk theory, each state in $\mathbb{S}_N$ is referred to as a vertex and the system is referred to as a quantum walker. We shall also refer to the elements of $\mathbb{S}_N$ as states, or sites.

\begin{figure}
\centerline{\includegraphics[width=0.4\textwidth]{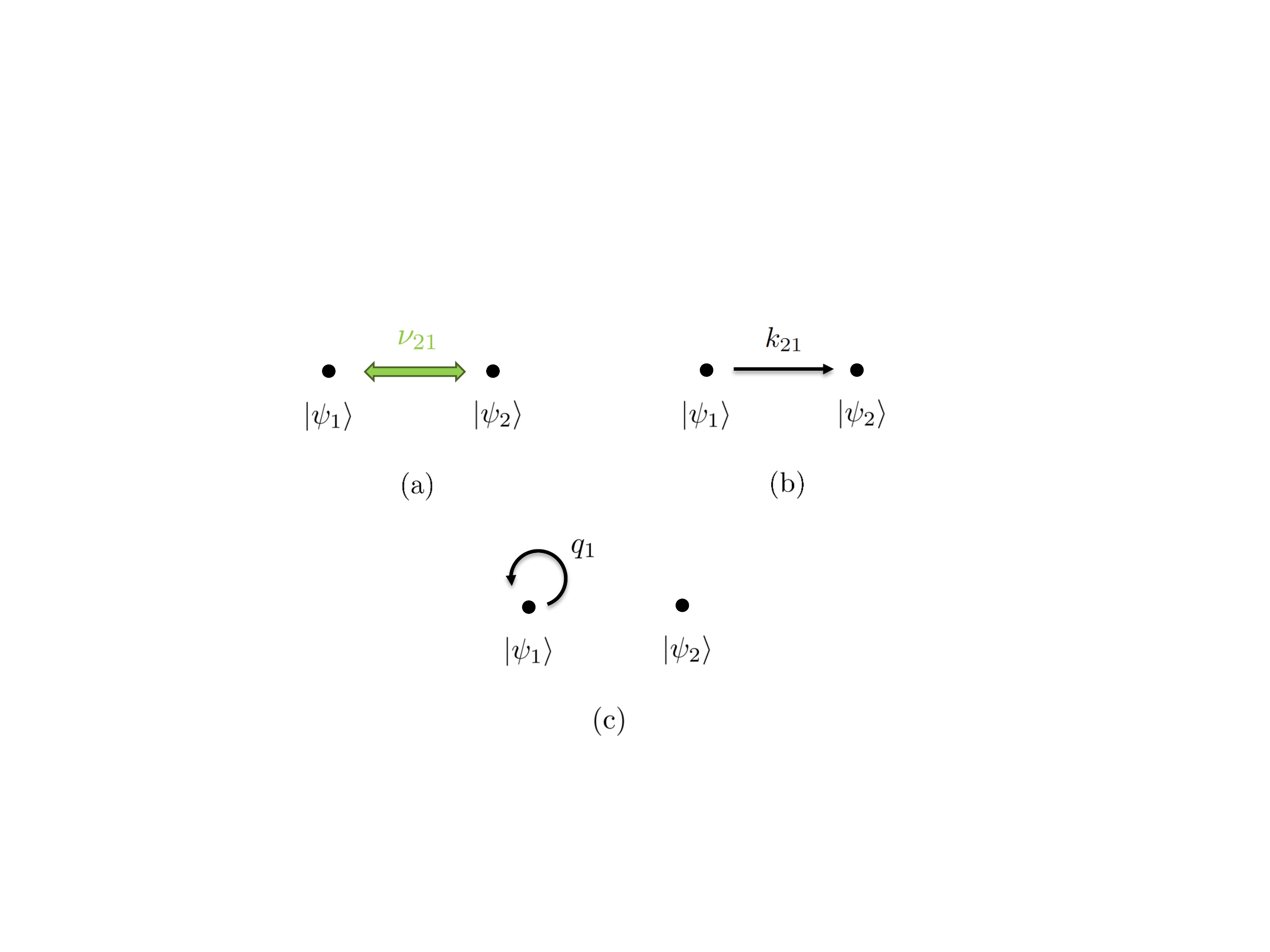}}
\caption{\label{BasicProcesses} Different types of edges considered in the quantum walk. (a) Coherent transition. (b) Incoherent transition. (c) Loop (dephasing).}
\end{figure}

We must therefore also define the exact manner in which the system makes transitions between different states (analogous to specifying a transtion matrix in classical Markov chains). We assume that our system is undergoing dynamics that can be composed from three basic processes. They are coherent transitions, incoherent population transfer, and dephasing. These processes are represented schematically in Fig.~\ref{BasicProcesses} for two vertices and are explained below.

\subsubsection{Coherent transition}
\label{[Hmn,Rho]}

The ability to be in a superposition of states is what differentiates the random walk of a quantum system from a classical one, so coherent evolution, i.e.~any evolution that adds coherences to our system, is indispensable in our theory. Coherent evolution between any two states $\ket{\psi_j}$ and $\ket{\psi_k}$ can be described by 
\begin{align}
\label{CoherentEvo}
	\frac{d}{dt} \, \rho(t) = -i \big[ \hat{H}_{jk},\rho(t) \big]  \;.
\end{align}
This is the familiar unitary evolution with a Hamiltonian given by
\begin{align}
	\hat{H}_{jk} = \omega_j \, \Qhat_j + \omega_k \, \Qhat_k + \Omega_{jk} \big( \Qhat_{jk} + \Qhat_{kj} \big)  \;,
\end{align}
where $\omega_k=\bra{\psi_k}\hat{H}_{jk}\ket{\psi_k}$ is the expectation of $\hat{H}_{jk}$ in the state $\ket{\psi_k}$, and
\begin{align}
	\hat{Q}_{jk} = \op{\psi_j}{\psi_k}, \quad \hat{Q}_{j} = \op{\psi_j}{\psi_j}.
\end{align}
We are assuming 
\begin{align}
	\Omega_{jk} = \Omega^*_{jk} = \Omega_{kj} \;,
\end{align}
so that $\hat{H}_{jk}$ is represented by a real and symmetric matrix. We can define the rate at which coherent evolution occurs by how quickly the transition probability varies in time under \eqref{CoherentEvo}. The transition probability for a system evolving according to \eqref{CoherentEvo} is
\begin{align}
	\alpha_{jk}(t) \equiv \big| \bra{\psi_j} \;\! e^{-i\hat{H}_{jk}t} \;\! \ket{\psi_k} \big|^2  \;.
\end{align}
It can be shown, see e.g.~\cite{CGKPK16}, that this evaluates to 
\begin{align}
\label{CoherentTransitionProb}
	\alpha_{jk}(t) = \frac{2\,\Omega^2_{jk}}{\nu^2_{jk}} \, \big[ 1 - \cos(\nu_{jk}\,t) \big]  \;,
\end{align}
where 
\begin{align}
\label{CoherentFreq}
	\nu_{jk} = \rt{(\omega_j-\omega_k)^2 + 4\Omega^2_{jk}}  \;.
\end{align}
An inspection of \eqref{CoherentTransitionProb} suggests that we should define the frequency of a coherent transition between any two states $\ket{\psi_j}$ and $\ket{\psi_k}$ to be $\nu_{jk}$. The symbol that we shall assign to coherent transitions is shown in Fig.~\ref{BasicProcesses}(a).

\subsubsection{Incoherent population transfer}
\label{DmnRho}

This is a process which transfers a fraction of the population in one state to another. It is shown in Fig.~\ref{BasicProcesses}(b) for the case of two states with the direction of the arrow indicating the direction of population transfer. For a system with $N$ states, the incoherent population transfer between any two states, say $\ket{\psi_m}$ and $\ket{\psi_n}$, occurring at a rate $k_{mn}$ is described by
\begin{align}
\label{IncohPopTrans}
	\frac{d}{dt} \, \rho(t) = k_{mn} \, \Ds_{mn} \, \rho(t) 
	                   \equiv k_{mn} \bigg[ \Qhat_{mn} \, \rho(t) \, \Qhat\dg_{mn} 
	                          - \frac{1}{2} \, \Qhat\dg_{mn} \, \Qhat_{mn} \, \rho(t) - \frac{1}{2} \, \rho(t) \, \Qhat\dg_{mn} \, \Qhat_{mn} \bigg].
\end{align}
Note the order of the indices defines the direction of the process so $\Ds_{mn} \ne \Ds_{nm}$ for $m \ne n$. Equation \eqref{IncohPopTrans} can be understood by considering an arbitrary state and propagating it over an infinitesimal interval $dt$. For simplicity let us consider the $N=2$ scenario. The result is then most apparent if we use the matrix representation in the basis $\mathbb{S}_2$. Assuming the process to occur with a rate $k_{21}$ and in the direction shown in Fig.~\ref{BasicProcesses}(b) we have
\begin{align}
\label{ADmatrix}
	\rho(t+dt) = \tbt{\rho_{11}(t)-k_{21}\,\rho_{11}(t)\,dt}{\rt{1-k_{21}\,dt}\,\rho_{12}(t)}
	                 {\rt{1-k_{21}\,dt}\,\rho_{21}(t)}{\rho_{22}(t)+k_{21}\,\rho_{11}(t)\,dt}  \;,
\end{align}
where the matrix elements of $\rho(t)$ are abbreviated by
\begin{align}
\label{RhoMatrixElements}
	\rho_{mn}(t) = \bra{\psi_m} \rho(t) \ket{\psi_n}  \;.
\end{align}
From \eqref{ADmatrix} we can see that a fraction of the population in $\ket{\psi_1}$ has been transferred to $\ket{\psi_2}$ and that this amount is given by $k_{21}\,\rho_{11}(t)\,dt$. We can interpret $\rho_{11}(t)$ as the prior probability of finding the system in state $\ket{\psi_1}$ at time $t$ and $k_{21}\,dt$ as the probability that the system will transition to $\ket{\psi_2}$ during $dt$ given that it is in $\ket{\psi_1}$ at time $t$. Notice that in transferring the population from $\ket{\psi_1}$ to $\ket{\psi_2}$ we also drive the system to a less coherent state. This is seen in \eqref{ADmatrix} as the off-diagonal terms of $\rho(t+dt)$ are less than the off-diagonal terms in $\rho(t)$ by a factor of $\rt{1-k_{21}\,dt}$. This is the sense in which we call the process modelled by \eqref{IncohPopTrans} incoherent population transfer. The usual rate-equation model of continuous-time Markov chains appears as a special case of incoherent population transfer with $\rho_{21}(t)=\rho_{12}(t)=0$.

\subsubsection{Loop (dephasing)}
\label{DephasingME}

We saw above that incoherent population transfer leads to a reduction of the system coherences. However, a system can also lose coherences without the simultaneous loss of populations. Such processes are called dephasing. In general we can control how classical (or quantum) a state $\ket{\psi_n}$ is by tuning its ability to share coherences with other states in the graph. Assuming a dephasing rate of $q_{n}$ for $\ket{\psi_n}$, the corresponding equation of motion for this process is
\begin{align}
\label{DephasingDefn}
	\frac{d}{dt} \, \rho(t) = q_{n} \, \Ds_{n} \, \rho(t) 
	                   \equiv q_{n} \bigg[ \Qhat_{n} \, \rho(t) \, \Qhat_{n} 
	                          - \frac{1}{2} \, \Qhat_{n} \, \rho(t) - \frac{1}{2} \, \rho(t) \, \Qhat_{n} \bigg]  \;.
\end{align}
Again, we illustrate this process for the simplest case of a two-state graph shown in Fig.~\ref{BasicProcesses}(c). Allowing state $\ket{\psi_1}$ to dephase with a rate of $q_1$, we can show that \eqref{DephasingDefn} changes only the system coherences by looking at the matrix representation of an arbitrary state in $\mathbb{S}_2$:
\begin{align}
\label{DephasingMatrix}
	\rho(t+dt) = \tbt{\rho_{11}(t)}{\rt{1-q_{1}\,dt}\,\rho_{12}(t)}
	                 {\rt{1-q_{1}\,dt}\,\rho_{21}(t)}{\rho_{22}(t)}  \;.
\end{align}
It is clear that \eqref{DephasingMatrix} has off-diagonal elements with the same form as the off-diagonal elements in \eqref{ADmatrix}, but now there is no transfer of populations. In fact, \eqref{DephasingDefn} is a special case of \eqref{IncohPopTrans} when $n=m$, and
\begin{align}
	k_{nn} \equiv q_n  \;.
\end{align}
That is, the population transfer is from state $\ket{\psi_n}$ back to itself so we expect the population of $\ket{\psi_n}$ to be unchanged. However, we saw above that incoherent population transfer also reduces the coherences in the graph so dephasing can be viewed simply as a loop shown in Fig.~\ref{BasicProcesses}(c).

\section{Distribution of hitting times}
\label{HTDdefn}

We are now in position to apply the quantum-jump prescription to calculate the distribution of hitting times. Consider now a graph where each state is connected to another by either one or all of the processes shown in Fig.~\ref{BasicProcesses}. The question that we would like to answer is how long it takes the system (or a quantum walker) to reach a particular state $\rhof \in \mathbb{S}_N$ for the first time assuming some initial state $\rhoi$. This length of time is called the hitting time. However, each time we run the quantum walk the time taken to reach $\rhof$ for the first time will be different, so the hitting time is actually a random variable. We denote the hitting time by $T(\rhof,\rhoi)$. What we would like to know is actually the distribution of hitting times if we were to run the quantum walk on the same graph many times. Since the hitting time here is a continuous variable, its distribution is governed by a probability density. The hitting probability density is a function $h(t;\rhof,\rhoi)$ such that
\begin{align}
\label{GeneralHTD}
	h(t;\rhof,\rhoi) \, dt
	= {\rm Pr} \Big\{ \rho(\tau) = \rhof \; \text{for} \; \tau \in [t,t+dt) \,, \, \rho(\tau) \ne \rhof \; \text{for} \; \tau \in (0,t)\; \Big| \; \rho(0) = \rhoi \Big\}  \;.
\end{align}
We are using the notation ${\rm Pr}\{ A | B \}$ for the probability of event $A$ occurring given that event $B$ has occurred and an equation like ${\rm Pr}\{\rho(t)=\rho'\}$ is to be read as the probability of finding the system in the state $\rho'$ at time $t$. The hitting distribution depends on the choice of $\rhoi$ and $\rhof$ for a given graph but once chosen they are fixed so $\rhoi$ and $\rhof$ are to be thought of as parameters in \eqref{GeneralHTD}. Without loss of generality we shall define the final state as the $N$th state in $\mathbb{S}_N$ as it is just a matter of labelling whereas the initial state will be left unspecified. That is
\begin{align}
	\rhof = \op{\psi_N}{\psi_N}=\hat{Q}_N  \;.
\end{align}
Since $\rhoi$ and $\rhof$ are fixed we will not write these out explicitly as arguments of $h$ unless it is helpful to do so.

\subsection{Derivation without dephasing}

We now derive a closed-form expression for the hitting time distribution on an arbitrary graph using quantum-jump formalism.
For simplicity we first consider a graph where there are no loops, i.e.~no dephasing:
\begin{align}
\label{GeneralGraphNoDeph}
	\frac{d}{dt} \, \rho(t) = \Ls \, \rho(t) 
	          \equiv - \frac{i}{2} \, \underset{m \ne n}{\sum_{m=1}^{N-1} \sum_{n=1}^{N-1}} \, \big[ \hat{H}_{mn},\rho(t) \big] 
	                 + \underset{m \ne n}{\sum_{m=1}^N \sum_{n=1}^N} \, k_{mn} \, \Ds_{mn} \, \rho(t)  \;.
\end{align}
Note that we have divided the sum over commutators by half to compensate for double counting. This is because $\hat{H}_{mn}=\hat{H}_{nm}$. In general not every state will be connected to every other state via all possible processes. When this is the case we can simply choose which processes to switch off in \eqref{GeneralGraphNoDeph} by setting the relevant rates zero. One may have also noticed that the sums over the coherent transitions (the commutator terms) do not include transitions to the final state $\ket{\psi_N}$. This is because only the incoherent transitions to $\ket{\psi_N}$ are detectable in the quantum-jump formalism. Later on (in Sec.~\ref{N+1Model}) we will show how the hitting probability density can be obtained when there are coherent transitions to $\ket{\psi_N}$ by introducing an artifical dissipative channel.

In order to apply the quantum-jump technique we first note that \eqref{GeneralHTD} can also be rewritten using Bayes's rule as
\begin{align}
\label{HittingBayes}
	h(t) \, dt
	= {}& {\rm Pr} \Big\{ \rho(\tau) = \hat{Q}_N \; \text{for} \; \tau \in [t,t+dt) \; \Big| \; \rho(\tau) \ne \hat{Q}_N \; \text{for} \; \tau \in (0,t) \,, \, \rho(0) = \rhoi \Big\}            \nn \\ & \times {\rm Pr} \Big\{ \rho(\tau) \ne \hat{Q}_N \; \text{for} \; \tau \in (0,t) \; \Big| \; \rho(0) = \rhoi \Big\}  \;.
\end{align}
Each factor on the right-hand side in \eqref{HittingBayes} can now be calculated using the prescription laid out in Sec.~\ref{MCW}. To do this we map \eqref{MultichannelL} directly to \eqref{GeneralGraphNoDeph} by making the following identification
\begin{gather}
\label{c=Q}
	\big\{ \hat{c}_q \big \}_q = \big\{ \rt{k_{mn}} \hat{Q}_{mn} \big\}_{m,n}  \;, \\
\label{H}
	\hat{H} = \frac{1}{2} \, \underset{m \ne n}{\sum_{m=1}^{N-1} \sum_{n=1}^{N-1}} \, \hat{H}_{mn}  \;.
\end{gather}
In terms of the multichannel theory of quantum jumps in Sec.~\ref{MultichannelQuantumJumps}, each channel is now labelled by an ordered pair $(m,n)$ where $m=1,2,\ldots N$, and $n=1,2,\ldots N$, and a quantum jump is simply the detection of a transition between any two nodes, say $\ket{\psi_m}$ and $\ket{\psi_n}$. The measurement efficiency for this quantum jump can then be labelled by $\eta_{mn}$, and is a number which corresponds to the fraction of jumps observed in the $\ket{\psi_n} \to \ket{\psi_m}$ incoherent transition. The actual number of jumps from $\ket{\psi_n}$ to $\ket{\psi_m}$ will always be greater than the observed number of jumps unless $\eta_{mn}=1$. We can then express the probability of finding the system in the final state $\ket{\psi_N}$ using the probability of detecting a jump to $\ket{\psi_N}$ provided that all transitions to $\ket{\psi_N}$ are monitored with unit efficiency. This identification then dictates how we should set $\eta_{mn}$ for all $m$ and $n$, namely that $\eta_{Nn}=1$ for any $n$. Transitions to states other than $\ket{\psi_N}$ are irrelevant so we simply do not monitor these transitions. Therefore we set $\eta_{mn}=0$ for every $m \ne N$. We must use this set of detection efficiencies in the formalism outlined in Sec.~\ref{MultichannelQuantumJumps} in order for it to be applicable for hitting-time calculations. This prompts us to assign a special label, $\bm{\eta}_\star$, for the detection efficiencies to be used:
\begin{align}
\label{EtaStar}
	\bm{\eta}_\star \, \Longleftrightarrow  \, \eta_{mn} = \delta_{mN}  \;.
\end{align}
The jump superoperator \eqref{JumpEta} on using \eqref{c=Q} and \eqref{EtaStar}, becomes 
\begin{align}
\label{Jstar}
	{\cal J}({\bm \eta}_\star) \, \rho \equiv \sum_{n=1}^{N-1}  \, k_{Nn} \, \hat{Q}_{Nn} \, \rho \; \hat{Q}\dg_{Nn}  \;.
\end{align}
An application of ${\cal J}({\bm \eta}_\star)$ effects an instantaneous collapse of the system to the state $\ket{\psi_N}$. From \eqref{MultichannelCountProb} and \eqref{c=Q}, the probability of detecting a jump in time $dt$ given $\rho(t)$ is thus
\begin{align}
\label{JumpProb}
	\wp_1(\bm{\eta}_\star;dt) = {\rm Tr}\big[ {\cal J}(\bm{\eta}_\star) \, \rho(t) \big] \, dt
                            = \sum_{n=1}^{N-1} \; k_{Nn} \, \bra{\psi_n} \rho(t) \ket{\psi_n} \, dt \;.
\end{align}
If $\rho(t)$ in \eqref{JumpProb} is a state where no transitions to $\ket{\psi_N}$ are observed in the time interval $[0,t)$ given some initial state $\rho(0)$, then \eqref{JumpProb} is exactly the first factor on the right-hand side of \eqref{HittingBayes}. Such a state can be directly obtained by solving the no-jump evolution defined by \eqref{MultichannelNoCountL}, \eqref{c=Q}, and \eqref{H}:
\begin{align}
\label{NoJumpEvolution}
	\frac{d}{dt} \, \bar{\rho}(t) \equiv \bar{\cal L}(\bm{\eta}_\star) \, \bar{\rho}(t) 
	= {}& - \frac{i}{2} \, \underset{m \ne n}{\sum_{m=1}^{N-1} \sum_{n=1}^{N-1}} \, \big[ \hat{H}_{mn},\bar{\rho}(t)\big] 
	      + \underset{m \ne n}{\sum_{m=1}^N \sum_{n=1}^N} \; k_{mn} \, (1-\delta_{mN}) \, \hat{Q}_{mn} \, \bar{\rho}(t) \, \hat{Q}\dg_{mn} \nn \\
	    & - \frac{1}{2} \, \underset{m \ne n}{\sum_{m=1}^N \sum_{n=1}^N} \; k_{mn} \, \big\{ \hat{Q}\dg_{mn} \hat{Q}_{mn}, \bar{\rho}(t) \big\}  \;.
\end{align}
It helps to rewrite $\bar{\cal L}(\bm{\eta}_\star)\;\! \bar{\rho}(t)$ so that its meaning is more obvious. Expanding the sum containing $(1-\delta_{mN})$ we get
\begin{align}
\label{EtaStarL}
	\bar{\cal L}(\bm{\eta}_\star) \, \bar{\rho}(t) 
	= {}& - \frac{i}{2} \, \underset{m \ne n}{\sum_{m=1}^{N-1} \sum_{n=1}^{N-1}} \, \big[ \hat{H}_{mn},\bar{\rho}(t)\big] 
	      + \underset{m \ne n}{\sum_{m=1}^N \sum_{n=1}^N} \; k_{mn} \, \hat{Q}_{mn} \, \bar{\rho}(t) \, \hat{Q}\dg_{mn}  \nn \\
	    & - {\cal J}({\bm \eta}_\star) \, \bar{\rho}(t) 
	      - \frac{1}{2} \, \underset{m \ne n}{\sum_{m=1}^N \sum_{n=1}^N} \; k_{mn} \, \big\{ \hat{Q}\dg_{mn} \hat{Q}_{mn}, \bar{\rho}(t) \big\} \;.
\end{align}
It is simple to see that
\begin{align}
\label{DoubleSumJumps}
	\underset{m \ne n}{\sum_{m=1}^N \sum_{n=1}^N} \; k_{mn} \, \hat{Q}_{mn} \, \bar{\rho}(t) \, \hat{Q}\dg_{mn}  
	= {\cal J}({\bm \eta}_\star) \, \bar{\rho}(t) + \underset{m \ne n}{\sum_{m=1}^{N-1} \sum_{n=1}^N} \; k_{mn} \, \hat{Q}_{mn} \, \bar{\rho}(t) \, \hat{Q}\dg_{mn}  \;,
\end{align}
and similarly
\begin{align}
\label{DoubleSumAnticomm}
	\underset{m \ne n}{\sum_{m=1}^N \sum_{n=1}^N} \; k_{mn} \, \big\{ \hat{Q}\dg_{mn} \hat{Q}_{mn}, \bar{\rho}(t) \big\}  
	= \sum_{n=1}^{N-1} \; k_{Nn} \, \big\{ \hat{Q}\dg_{Nn} \hat{Q}_{Nn}, \bar{\rho}(t) \big\}
	  + \underset{m \ne n}{\sum_{m=1}^{N-1} \sum_{n=1}^N} \; k_{mn} \, \big\{ \hat{Q}\dg_{mn} \hat{Q}_{mn}, \bar{\rho}(t) \big\} \;.
\end{align}
Substituting \eqref{DoubleSumJumps} and \eqref{DoubleSumAnticomm} into \eqref{EtaStarL} gives
\begin{align}
\label{SimpleFormLbar}
	\bar{\cal L}(\bm{\eta}_\star) \, \bar{\rho}(t) 
	={}& - \frac{i}{2} \, \underset{m \ne n}{\sum_{m=1}^{N-1} \sum_{n=1}^{N-1}} \, \big[ \hat{H}_{mn},\bar{\rho}(t)\big]  
	     + \underset{m \ne n}{\sum_{m=1}^{N-1} \sum_{n=1}^N} \; k_{mn} \, \Ds_{mn} \, \bar{\rho}(t)  
	     + \sum_{n=1}^{N-1} \; k_{Nn} \, \bar{\Ds}_{Nn} \, \bar{\rho}(t)  \;,
\end{align}
where we have defined 
\begin{align}
\label{DbarDefn}
	\bar{\Ds}_{Nn} \, \rho = - \frac{1}{2} \, \big\{ \hat{Q}\dg_{Nn} \hat{Q}_{Nn}, \rho \big\}  \;.
\end{align}
As already described at the outset, the first term in \eqref{SimpleFormLbar} describes coherent population transfer between states other than the final state $\ket{\psi_N}$ (recall Sec.~\ref{[Hmn,Rho]}). The second term in \eqref{SimpleFormLbar} describes incoherent transitions that either move populations out of $\ket{\psi_N}$, or between any two states except for $\ket{\psi_N}$ (see Sec.~\ref{DmnRho}). The last term in \eqref{SimpleFormLbar} is similar in form to $\Ds_{mn}$ except that terms of the form $\hat{Q}_{Nn} \, \bar{\rho}(t) \, \hat{Q}\dg_{Nn}$ have been removed. But these are the terms which describe jumps to the final state. Therefore $\bar{\Ds}_{Nn}$ is a superoperator which describes how the system evolves conditioned on no jumps to $\ket{\psi_N}$. Substituting the formal solution of \eqref{NoJumpEvolution} into \eqref{JumpProb} we thus obtain 
\begin{align}
	{}& {\rm Pr} \Big\{ \rho(\tau) = \hat{Q}_N \; \text{for} \; \tau \in [t,t+dt) \; \Big| \; \rho(\tau) \ne \hat{Q}_N \; \text{for} \; \tau \in (0,t) \,, \, \rho(0) = \rhoi \Big\}  
			\nn \\
\label{ConditionedHTD}
	{}& = \sum_{n=1}^{N} \; k_{Nn} \, 
	      \frac{\bra{\psi_n} \;\! e^{\bar{\Ls}(\bm{\eta}_\star)\,t} \rhoi \;\! \ket{\psi_n}}{{\rm Tr}\big[ e^{\bar{\Ls}(\bm{\eta}_\star)\,t} \rhoi \big]}  \; dt \;.
\end{align}
Note that \eqref{NoJumpEvolution} produces a trace-decreasing state, which is why we have used an overbar for the density operator and the generator of time evolution. Following Sec.~\ref{MultichannelQuantumJumps} we see that the norm of $\bar{\rho}(t)$ is precisely the probability of no jumps in the interval $[0,t)$, given some initial state i.e.
\begin{align}
\label{NoJumpProbability}
	{\rm Pr} \Big\{ \rho(\tau) \ne \hat{Q}_N \; \text{for} \; \tau \in [0,t) \; \Big| \; \rho(0) = \rhoi \Big\} = {\rm Tr}\big[ e^{\bar{\Ls}(\bm{\eta}_\star)\,t} \rhoi \;\! \big]  \;.
\end{align}
The product of \eqref{ConditionedHTD} and \eqref{NoJumpProbability} simply cancels the norm of $\bar{\rho}(t)$ giving the final expression for the hitting probability density as
\begin{align}
\label{FinalHTD}
	h\big( t;\hat{Q}_N,\rhoi \big) = \sum_{n=1}^{N-1} \; k_{Nn} \, \bra{\psi_n} \, e^{\bar{\Ls}(\bm{\eta}_\star)\,t} \, \rhoi \;\! \ket{\psi_n}  \;.
\end{align}
This completes our derivation of the hitting distribution using the quantum-jump method.

\subsection{Examples}
\begin{figure}[t]
\centerline{\includegraphics[width=0.8\textwidth]{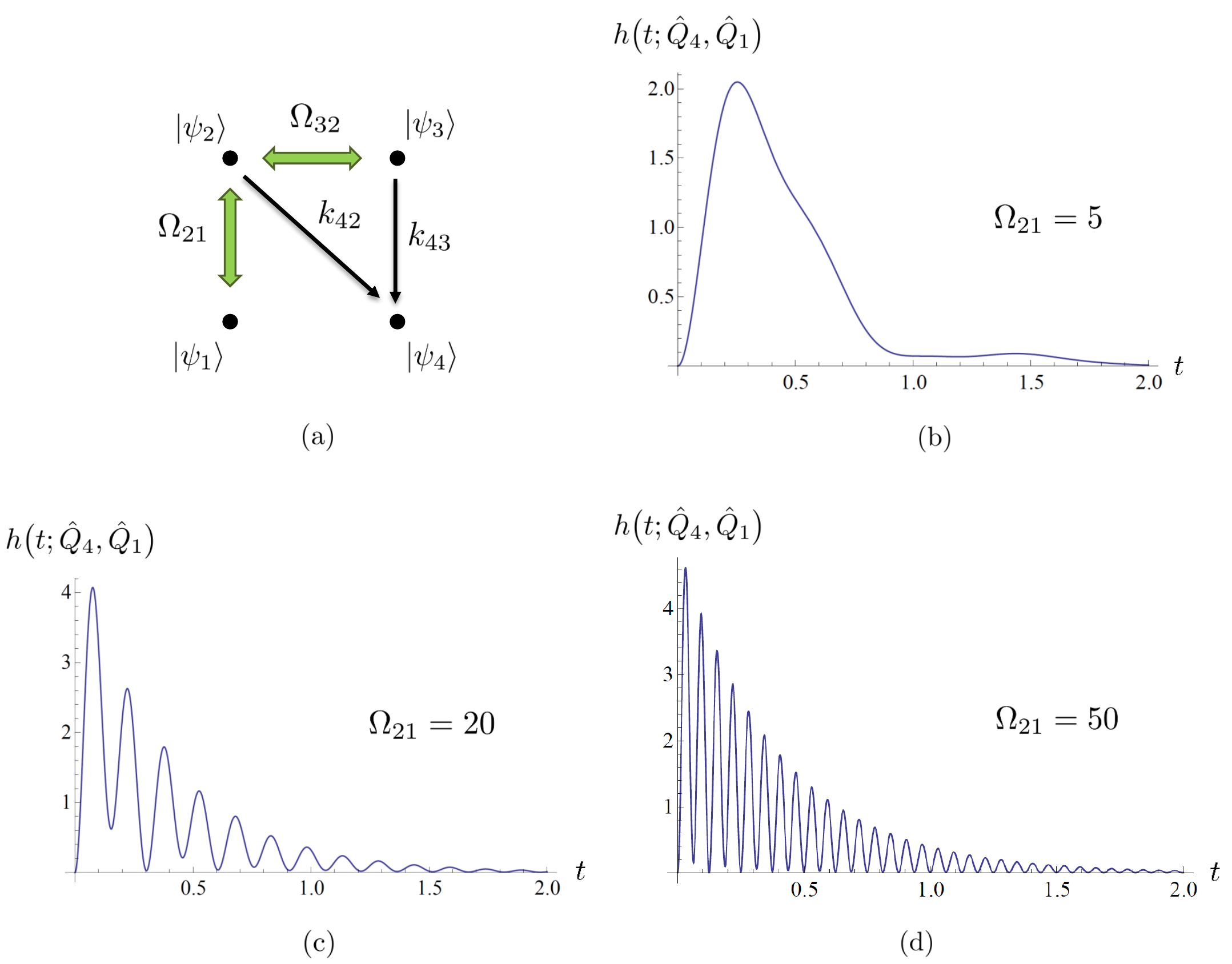}}
\caption{\label{HTDwithoutDephasingA} Recall from Sec.~\ref{[Hmn,Rho]} that a Hamiltonian, say $\hat{H}_{21}$ is parameterised by the three numbers $\omega_1$, $\omega_2$, and $\Omega_{21}$, which correspond to the diagonal and off-diagonal matrix elements of $\hat{H}_{21}$ in the site basis. Here we set $\omega_1=1$, $\omega_2=3$, and $\omega_3=5$ which are kept constant throughout. We then set the transition rates to be $\Omega_{32}=k_{42}=k_{43}=5$ and change only $\Omega_{21}$ whose value is shown in the plot. Note that $| \omega_1-\omega_2 | \ll 2 \Omega_{21}$, so according to \eqref{CoherentFreq} of Sec.~\ref{[Hmn,Rho]} $\nu_{21} \approx 2\Omega_{21}$.}
\end{figure}
\begin{figure}[t]
\centerline{\includegraphics[width=0.9\textwidth]{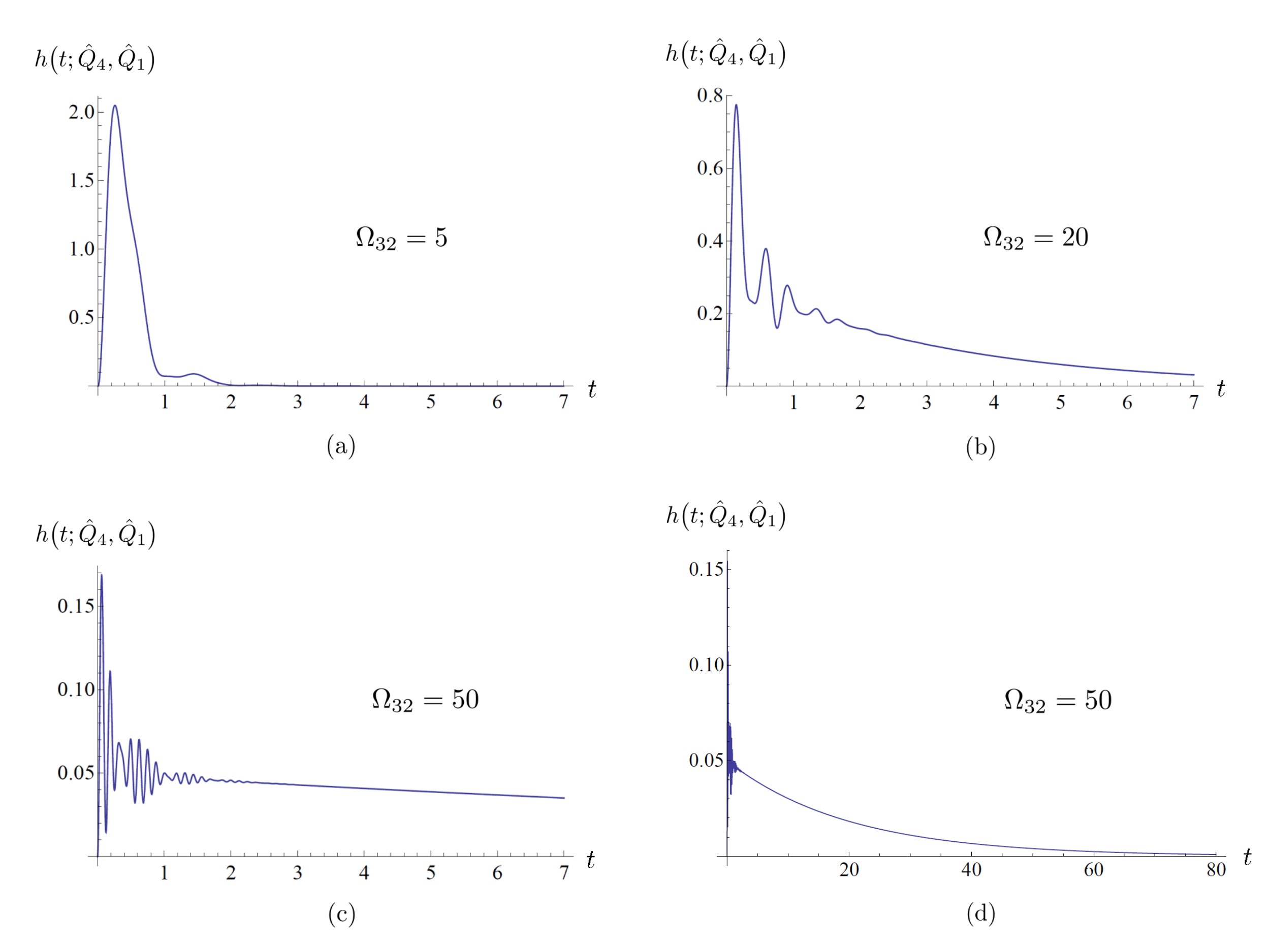}}
\caption{\label{HTDwithoutDephasingB} Hitting distributions for various values of $\Omega_{32}$ (shown in the plots). We fix $\Omega_{21}=5$ while keeping the remaining parameters the same as in Fig.~\ref{HTDwithoutDephasingA} ($k_{42}=k_{43}=5$ and $\omega_1=1$, $\omega_2=3$, $\omega_3=5$). Note that (c) and (d) are for identical parameters just that (c) plots the hitting distribution for short times.}
\end{figure}

We now illustrate \eqref{FinalHTD} with a simple example. Consider the graph shown in Fig.~\ref{HTDwithoutDephasingA}(a). This is an example with $N=4$ so by our convention the final state is $\rhof=\op{\psi_4}{\psi_4}$. We will let the initial state be $\rhoi=\op{\psi_1}{\psi_1}$. In Fig.~\ref{HTDwithoutDephasingA}(b), (c), and (d), we plot the hitting distribution for various amounts of coherence between states $\ket{\psi_1}$ and $\ket{\psi_2}$ while keeping other parameters constant (see figure caption for their values). We observe that as  $\Omega_{21}$ is increased from 5 [Fig.~\ref{HTDwithoutDephasingA}(b)] to 50 [Fig.~\ref{HTDwithoutDephasingA}(d)], the hitting distribution starts to show more and more oscillations due to the increased level of coherence. Classically, only incoherent transitions are permitted so hitting distributions do not exhibit oscillations. We can see from Fig.~\ref{HTDwithoutDephasingA}(b) that for $\Omega_{21}=5$ the oscillations in the hitting distribution are dying out and the quantum walk is in transition to the classical regime. Furthermore, the oscillations in the hitting distribution can actually reach zero (or near zero) at its minimum meaning that there are times at which it is impossible to find the walker at $\ket{\psi_4}$. Again, such features would not appear in a purely classical walk since incoherent transitions cannot produce any interference effects. For variation we show in Fig.~\ref{HTDwithoutDephasingB} what happens when we change $\Omega_{32}$ instead of $\Omega_{21}$ while keeping every other parameter constant (see the figure caption for the actual values). Like Fig.~\ref{HTDwithoutDephasingA}, the amount of oscillations increases as $\Omega_{32}$ is increased, but now this is accompanied by a lengthening of the tail of the hitting distribution. The shape of the hitting distribution depends on the exact manner in which the probability amplitudes for the different paths of the quantum walk interfere. A comparison of Fig.~\ref{HTDwithoutDephasingA} with Fig.~\ref{HTDwithoutDephasingB} illustrates that hitting times of a given graph can be much more sensitive to one coherent transition but not others. We will have more to say about this in Sec.~\ref{AvgHittingTime}.

\subsection{Including dephasing}
\label{IncludingDephasing}

We obtained \eqref{FinalHTD} based on \eqref{GeneralGraphNoDeph} which does not have the dephasing processes defined in \eqref{DephasingDefn}. The omission of dephasing connections was simply to make the derivation easier to follow. If we include dephasing in $\Ls$ then the dephasing terms will simply carry through to $\bar{\Ls}(\bm{\eta}_\star)$. The expression for the hitting-time probability density will still be given by \eqref{FinalHTD}, just with an $\bar{\Ls}(\bm{\eta}_\star)$ that includes dephasing connections. Stated more formally, we can consider a graph defined by 
\begin{align}
\label{GeneralGraphWithDeph}
	\frac{d}{dt} \, \rho(t) = \Ls \, \rho(t) 
	                   \equiv - \frac{i}{2} \, \underset{m \ne n}{\sum_{m=1}^{N-1} \sum_{n=1}^{N-1}} \, \big[ \hat{H}_{mn},\rho(t) \big] 
	                          + \underset{m \ne n}{\sum_{m=1}^N \sum_{n=1}^N} \; k_{mn} \, \Ds_{mn} \, \rho(t)  
	                          + \sum_{n=1}^N \; q_{n} \, \Ds_{n} \, \rho(t)  \;.
\end{align}
Note that the total number of incoherent channels in this case is given by $L=N^2$ and we can make the following identification for the Lindblad operators in \eqref{MultichannelL}:
\begin{gather}
\label{DephasingHatc}
	\big\{ \hat{c}_q \big \}_{q=1}^{N^2-N} = \big\{ \rt{k_{mn}} \, \hat{Q}_{mn} \big\}_{m,n}  \;, \quad	
	\big\{ \hat{c}_q \big \}_{q=N^2-N+1}^{N^2} = \big\{ \rt{q_{n}} \, \hat{Q}_{n} \big\}_{n}  \;.
\end{gather}
The Hamiltonian in \eqref{MultichannelL} is still given by \eqref{H}. The detection efficiencies should now be set according to  
\begin{align}
	\bm{\eta}_\star \, \Longleftrightarrow \, \eta_{mn}=\delta_{mN} \;, \quad  n=1,2,\ldots,L/2 \;.
\end{align}
This then gives
\begin{align}
\label{SimpleFormLbarWithDeph}
	\bar{\cal L}(\bm{\eta}_\star) \, \bar{\rho}(t) 
	= {}& - \frac{i}{2} \, \underset{m \ne n}{\sum_{m=1}^{N-1} \sum_{n=1}^{N-1}} \, \big[ \hat{H}_{mn},\bar{\rho}(t)\big]  
	      + \sum_{n=1}^N \; q_{n} \, \Ds_{n} \, \bar{\rho}(t)   \nn \\
	    & + \underset{m \ne n}{\sum_{m=1}^{N-1} \sum_{n=1}^N} \; k_{mn} \, \Ds_{mn} \, \bar{\rho}(t)  
	      + \sum_{n=1}^{N-1} \; k_{Nn} \, \bar{\Ds}_{Nn} \, \bar{\rho}(t)  \;. 
\end{align}
Replacing $\bar{\Ls}(\bm{\eta}_\star)$ in \eqref{FinalHTD} with \eqref{SimpleFormLbarWithDeph} then gives the distribution of hitting times for the general graph.
\begin{figure}[t]
\centerline{\includegraphics[width=0.8\textwidth]{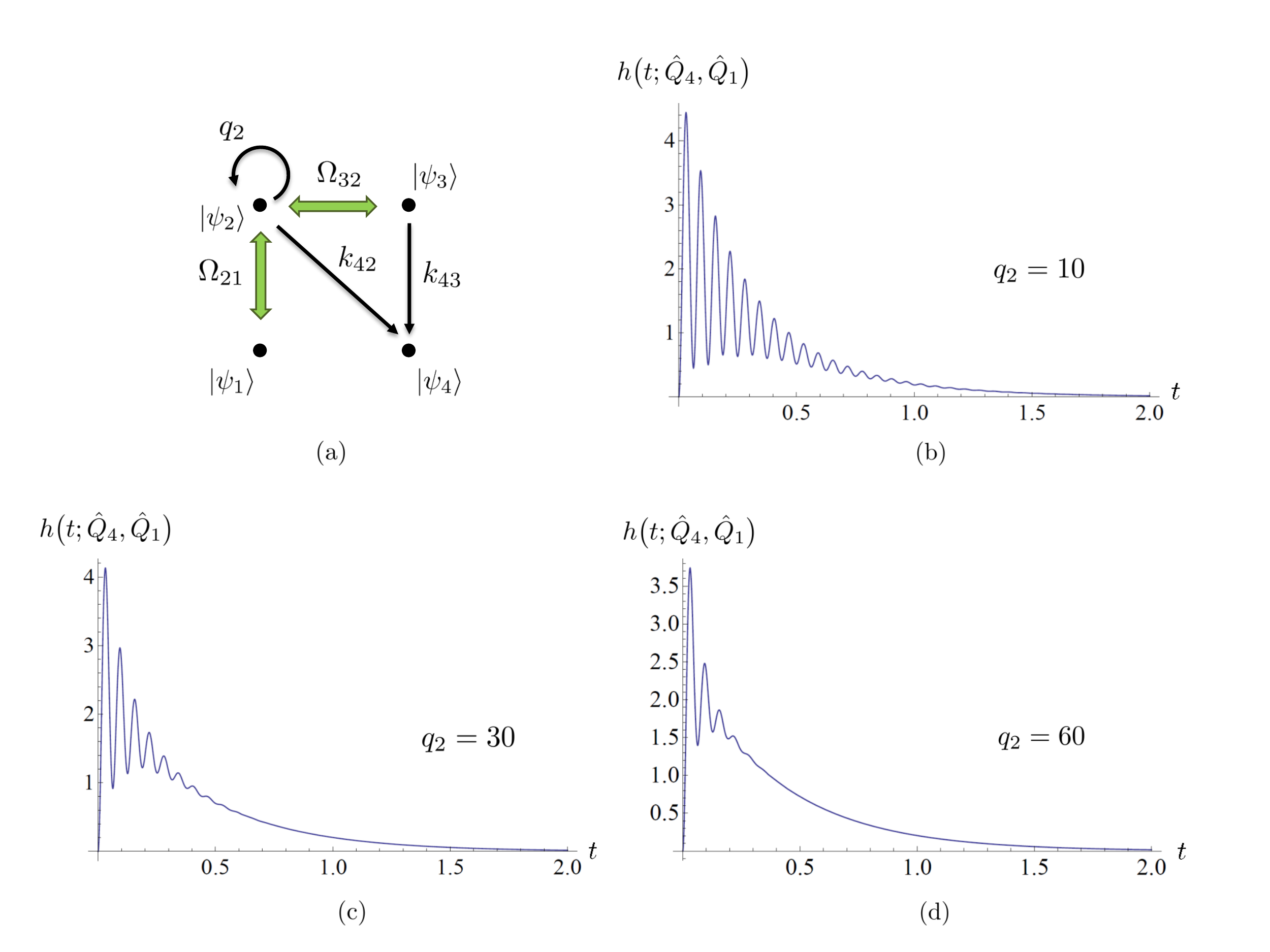}}
\caption{\label{HTDwithDephasing} $\Omega_{32}=k_{42}=k_{43}=5$, $\Omega_{21}=50$, $\omega_1=1$, $\omega_2=3$, and $\omega_3=5$. See Fig.~\ref{HTDwithoutDephasingA}(d) for comparison.}
\end{figure}

The reason why adding dephasing connections to the graph does not change the form of hitting distribution as a function of $\bar{\Ls}(\bm{\eta}_\star)$ is because the quantum-jump method works by detecting transitions to the final state $\ket{\psi_N}$. Of the three types of connections that we consider, only two can induce transitions between states---coherent and incoherent population transfer. The third type, i.e~dephasing, changes only the coherences in the system. Of course changing coherences in the graph does affect how quickly the final state is reached. That is to say, the distribution of hitting times for a state diagram with dephasing will be a different function of $t$ compared to one without dephasing but otherwise identical in all respects.

For illustration we include in Fig.~\ref{HTDwithDephasing} hitting distributions for the graph with dephasing of $\ket{\psi_2}$, see Fig.~\ref{HTDwithDephasing}(a). The plots in Fig.~\ref{HTDwithDephasing} are identical to Fig.~\ref{HTDwithoutDephasingA}(d) except with various amounts of dephasing added as indicated in the plots. Here the dephasing rate is $q_2$ and as we increase this rate the quantum walk becomes more classical. This can be seen in passing from Fig.~\ref{HTDwithDephasing}(b) to (d) where the oscillations which we attributed to the quantum-mechanical nature of the walk in Fig.~\ref{HTDwithoutDephasingA} gradually die out. This is to be expected since by increasing $q_2$ we are making the walk more classical, in other words, making it more and more difficult to share coherences between $\ket{\psi_2}$ and other nodes in the graph. Next we calculate the moments of $h\big( t;\hat{Q}_N,\rhoi \big)$.

\section{Hitting statistics}
\label{AvgHittingTime}

\subsection{The $n$th moment}
\label{nthMomentResult}

The $n$th raw statistical moment of the hitting time $T(\hat{Q}_N,\rhoi)$ is defined as
\begin{align}
\label{nthMomentDefn}
	{\rm E}\big[T^n(\hat{Q}_N,\rhoi)\big] = \int_0^\infty  dt \; t^n \; h(t;\hat{Q}_N,\rhoi)  \;,
\end{align}
where we are using the notation ${\rm E}[X]$ to mean the ensemble average of an arbitrary random variable $X$. Although having the hitting-time distribution is formally equivalent to knowing the statistics of $T(\hat{Q}_N,\rhoi)$, deriving a closed-form expression for \eqref{nthMomentDefn} is still a nontrivial task. We will show in Sec.~\ref{nthMomentProof} that \eqref{nthMomentDefn} evaluates to 
\begin{align}
\label{M(x)Final}
	{\rm E}\big[T^n(\hat{Q}_N,\rhoi)\big] = (-1)^{n+1} \, n! \sum_{m=1}^{N-1} \, k_{Nm} \, \bra{\psi_m} \, \Lstar^+{}^{(n+1)}  \, \rhoi \;\! \ket{\psi_m} 
\end{align}
where we have defined the shorthand
\begin{align}
	\Lstar = \bar{\cal L}(\bm{\eta}_\star)  \;,
\end{align}
and $\Lstar^+$ is the Moore--Penrose pseudoinverse of $\Lstar$, defined by
\begin{align}
\label{Lstar+Defn}
	\Lstar^+\, \Lstar \, \Lstar^+ = \Lstar^+  \;, \quad
	\Lstar \, \Lstar^+ \, \Lstar = \Lstar  \;, \quad
	\big[ \Lstar^+ \, \Lstar \big]\dg = \Lstar^+ \, \Lstar \;, \quad
	\big[ \Lstar \, \Lstar^+ \big]\dg = \Lstar \, \Lstar^+ \;. 
\end{align}
Note the superoperator Hermitian conjugate in \eqref{Lstar+Defn} is defined with respect to the Hilbert--Schmidt inner product ${\rm Tr}[\hat{X}\dg\hat{Y}]$ between any two bounded operators $\hat{X}$ and $\hat{Y}$. We say an arbitrary superoperator ${\cal A}$ is Hermitian, i.e.~${\cal A}={\cal A}\dg$ if and only if 
\begin{align}
	\Big( {\rm Tr}\big[\hat{X}\dg {\cal A} \hat{Y} \big] \Big)^* = {\rm Tr}\big[\hat{Y}\dg {\cal A} \hat{X} \big]  \quad \forall \; \hat{X} \,, \hat{Y}  \;.
\end{align}
The Moore--Penrose pseudoinverse always exists, is unique, and recovers the standard inverse when $\Lstar$ is invertible. One may wonder how $\Lstar^+$ can actually be computed for a given graph. To find the pseudoinverse it will be most convenient to use a matrix representation for $\Lstar$. This can be accomplished by using $\{\hat{Q}_{mn}\}_{m,n}$ as an operator basis, in which case the Hilbert--Schmidt inner product allows us to represent an arbitrary state $\rho$ by a vector $\bm{\rho}$ whose elements are $\rho_{mn}$. Note in this representation each row of $\bm{\rho}$ is labelled by two indices so $\bm{\rho}$ is a $N^2 \times 1$ vector. Similarly $\Lstar$ can be represented by an $N^2 \times N^2$ matrix $(\Lstar)$ whose elements are
\begin{align}
\label{MatRepLstar}
	(\Lstar)_{jk,mn} = {\rm Tr}\big[ \hat{Q}\dg_{jk} \Lstar \hat{Q}_{mn} \big] \;.
\end{align}
The superoperator pseudoinverse $\Lstar^+$ will then be faithfully represented by the Moore--Penrose pseudoinverse of $(\Lstar)$, i.e.~$(\Lstar^+)=(\Lstar)^+$. Equation \eqref{MatRepLstar} will be useful in Sec.~\ref{Examples} when we consider simple examples. There, we will consider one example where $\Lstar^{-1}$ does not exist and another where it does. This will explicitly illustrate the use of the Moore--Penrose pseudoinverse in \eqref{M(x)Final}.

The two most considered hitting statistics are the average and variance. We therefore pay special attention to these two quantities. On setting $n=1$ in \eqref{M(x)Final} we obtain the average hitting time:
\begin{align}
\label{E[t]MPInv}
	{\rm E}[\;\!T(\hat{Q}_N,\rhoi)\;\!] = \sum_{m=1}^{N-1} \, k_{Nm} \, \bra{\psi_m} \, \Lstar^+{}^2  \, \rhoi \;\! \ket{\psi_m}  \;.
\end{align}
Setting $n=2$ in \eqref{M(x)Final} and using the result for the average we arrive at an expression for the variance
\begin{align}
\label{V[T]Final}
	{\rm V}\big[T(\hat{Q}_N,\rhoi)\big]  
	= - 2 \sum_{m=1}^{N-1} \, k_{Nm} \, \bra{\psi_m} \, \Lstar^+{}^3 \rhoi \ket{\psi_m} 
	  -  \Bigg( \sum_{m=1}^{N-1} \, k_{Nm} \, \bra{\psi_m} \, \Lstar^+{}^2  \, \rhoi \;\! \ket{\psi_m} \Bigg)^2 \;. 
\end{align}
Our expression for the $n$th moment makes the evaluation of the hitting statistics a simple and efficient procedure given $\Lstar$. For example, evaluating the average hitting time by setting $n=1$ in \eqref{nthMomentDefn} and evaluating the resulting integral requires the determination of an optimal upper limit to truncate the integral in order to balance the amount of simulation time and numerical precision. In particular, when $h(t)$ has a long tail, the calculation of the average hitting time via \eqref{nthMomentDefn} can be cumbersome. Equation \eqref{E[t]MPInv} has the advantage that it avoids these issues.

\subsection{Proof of the $n$th moment}
\label{nthMomentProof}

To derive \eqref{M(x)Final} we consider the moment-generating function $M_T(x)$ defined by
\begin{align}
\label{M(x)def}
	M_T(x) = {\rm E}\big[ e^{xT} \big]   \;.
\end{align}
Given $M_T(x)$, all higher-order moments of $T$ can be calculated as
\begin{align}
\label{nthMoment}
	{\rm E}\big[ T^n \big] = \bigg[ \frac{d^n}{dx^n} \; M_T(x) \bigg|_{x=0}    \;.
\end{align}
A formula for $M_T(x)$ is simple to derive by substituting the expression for $h(t;\hat{Q}_N,\rhoi)$ from \eqref{FinalHTD} into \eqref{M(x)def}. We obtain
\begin{align}
	M_T(x) = {}& \sum_{m=1}^{N-1} \; k_{Nm} \, \bra{\psi_m} \int_0^\infty dt \; e^{(x \mathbbm{1} + \Lstar)t} \, \rhoi \, \ket{\psi_m}   \\
				 = {}& \sum_{m=1}^{N-1} \; k_{Nm} \, \bra{\psi_m} \,
				       \big( x \mathbbm{1} + \Lstar \big)^{-1} \Big[ e^{(x \mathbbm{1} + \Lstar)t} \Big|_0^\infty \, \rhoi \, \ket{\psi_m}  \\
\label{limM(x)}
         = {}& \sum_{m=1}^{N-1} \; k_{Nm} \, \bra{\psi_m} \, \big( x \mathbbm{1} + \Lstar \big)^{-1} \,
				       \Big( \lim_{t\to\infty} e^{xt} \, e^{\Lstar t} - \mathbbm{1} \Big) \rhoi \, \ket{\psi_m} \;.
\end{align}
Here we are using $\mathbbm{1}$ for the superoperator identity. We now need to evaluate the $t \longrightarrow \infty$ limit in \eqref{limM(x)}. To do so we first note that 
\begin{align}
\label{ZeroSteadyState}
	\lim_{t \to \infty} e^{\Lstar t} \rhoi  = 0  \;.
\end{align}
This simply says that the steady state for the dynamics defined by the generator $\Lstar$ is the zero state (represented by a matrix with all elements equal to zero). We can prove this by showing that \eqref{SimpleFormLbar} produces a trace-decreasing density operator. It is simple to check from \eqref{IncohPopTrans} that $\Ds_{mn}$ is a traceless superoperator:
\begin{align}
	{\rm Tr}\big[ \Ds_{mn} \;\!\rho  \big] = 0  \;, \; \forall \; m,n \;.
\end{align}
It is also simple to see that the trace of a commutator is always zero. Therefore taking the trace of \eqref{SimpleFormLbar} we get
\begin{align}
	\frac{d}{dt} \, {\rm Tr}\big[ \bar{\rho}(t) \big] = {}& \sum_{m=1}^{N-1} \, k_{Nm} \, {\rm Tr}\big[ \bar{\Ds}_{Nm} \;\! \bar{\rho}(t) \big] \\
                                                    = {}& - \frac{1}{2} \, \sum_{m=1}^{N-1} \, k_{Nm} \, {\rm Tr}\big[ \hat{Q}\dg_{Nm} \, \hat{Q}_{Nm} \, \bar{\rho}(t) \big]  \;.
\end{align}
The traces on the right-hand side are positive since they are averages of positive operators in the state $\bar{\rho}(t)$. They are in fact just the diagonal elements of the unnormalised density operator $\bar{\rho}(t)$. Since the rates $k_{Nm}$ must also be positive for every value of $m$, ${\rm Tr}[\bar{\rho}(t)]$ has a negative rate of change, which means that $\bar{\rho}(t)$ will eventually decay to zero. We now assume the evolution under $\exp(\Lstar t)$ dominates over $\exp(xt)$ [i.e.~the approach to zero under $\exp(\Lstar t)$ is faster than the rise of $\exp(xt)$] then the $t\longrightarrow\infty$ limit can be dropped to give
\begin{align}
	M_T(x) =  - \sum_{m=1}^{N-1} \; k_{Nm} \, \bra{\psi_m} \, \big( x \mathbbm{1} + \Lstar \big)^{-1} \rhoi \, \ket{\psi_m} \;.
\end{align}
It is easy to see that 
\begin{align}
	\frac{d^n}{dx^n} \, \big( x \mathbbm{1} + \Lstar \big)^{-1} = (-1)^n \, n! \, \big( x \mathbbm{1} + \Lstar \big)^{-(n+1)}  \;.
\end{align}
Using \eqref{nthMoment} we therefore arrive at
\begin{align}
\label{E[Tn]Linv}
	{\rm E}\big[T^n(\hat{Q}_N,\rhoi)\big] = (-1)^{n+1} \, n! \sum_{m=1}^{N-1} \, k_{Nm} \, \bra{\psi_m} \, \Lstar^{-(n+1)}  \, \rhoi \;\! \ket{\psi_m}  \;.
\end{align}
We assumed that $\Lstar$ has an inverse in our proof and it is natural to ask what if $\Lstar^{-1}$ does not exist? If the inverse of $\Lstar$ does not exist then one can replace $\Lstar^{-1}$ by $\Lstar^+$, the Moore--Penrose pseudoinverse. As mentioned in Sec.~\ref{nthMomentResult}, $\Lstar^+$ always exists, is unique, and coincides with $\Lstar^{-1}$ if it can be found. The replacement of $\Lstar^{-1}$ by $\Lstar^+$ in \eqref{E[Tn]Linv} then gives \eqref{M(x)Final}. Equations \eqref{E[t]MPInv} and \eqref{V[T]Final} are obtained from \eqref{M(x)Final} which we have derived using the moment-generating function but they should also be derivable by directly doing the integral in \eqref{nthMomentDefn} with $n=1$ and $n=2$ respectively. We show that this is indeed possible in Appendices~\ref{AppA} and \ref{AppB}, which serve as independent proofs for the mean and variance of the hitting time. Let us now illustrate how these results work with a few simple examples below. We consider only the mean and variance of $T$.

\subsection{Examples}
\label{Examples}

\subsubsection{Example 1: An absorbing state for $N=2$}

\begin{figure}[t]
\centerline{\includegraphics[width=0.6\textwidth]{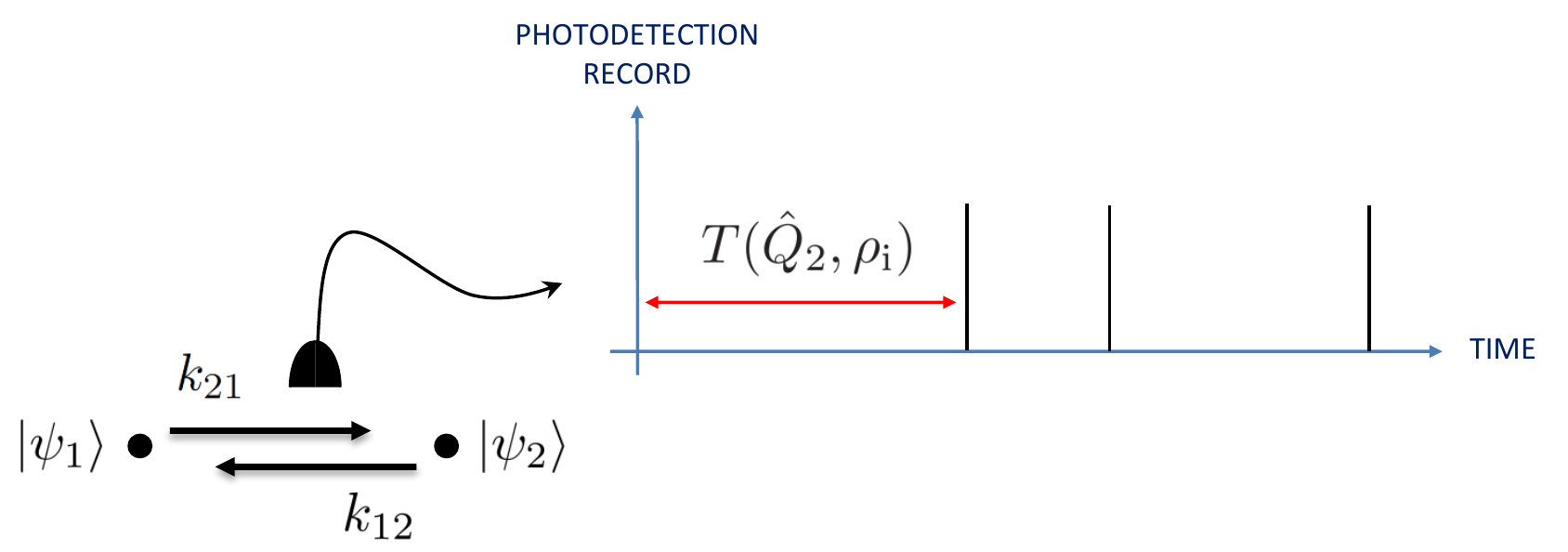}}
\caption{\label{TwoStateExample} A simple example with $N=2$. We are concerned with the hitting time for state $\ket{\psi_2}$ so we are only detecting the $\ket{\psi_1} \longrightarrow \ket{\psi_2}$ transition.}
\end{figure}

For our first example we consider the case of a simple incoherent transition between two states shown in Fig.~\ref{TwoStateExample} with $k_{12}=0$. It is clear that $\ket{\psi_2}$ is an absorbing state and we will see that our hitting statistics says so as well. The graph is defined by
\begin{align}
	{\cal L} \, \rho(t) = k_{21} \bigg[ \, \Qhat_{21} \, \rho(t) \, \Qhat\dg_{21} - \frac{1}{2} \, \Qhat_{1} \, \rho(t) - \frac{1}{2} \, \rho(t) \, \Qhat_{1} \, \bigg]  \;,
\end{align}
where we have used $\hat{Q}\dg_{21}\hat{Q}_{21}=\hat{Q}_1$. The average and variance of $T(\hat{Q}_2,\rhoi)$ according to \eqref{E[t]MPInv} and \eqref{V[T]Final} are thus
\begin{gather}
\label{TwoStateE[T]}
	{\rm E}[\;\!T(\hat{Q}_2,\rhoi)\;\!] = k_{21} \, \bra{\psi_1} \, \Lstar^+{}^2  \, \rhoi \, \ket{\psi_1} \;,  \\
\label{TwoStateV[T]}
	{\rm V}[\;\!T(\hat{Q}_2,\rhoi)\;\!] = - 2 \, k_{21} \, \bra{\psi_1} \, \Lstar^+{}^3  \, \rhoi \, \ket{\psi_1} 
	                                      - \big( k_{21} \, \bra{\psi_1} \, \Lstar^+{}^2  \, \rhoi \, \ket{\psi_1} \big)^2  \;.
\end{gather}
This then gives $\Lstar$ as 
\begin{align}
\label{TwoStateLbar1}
	\Lstar \, \rho(t) = - \frac{k_{21}}{2} \, \bigg[ \, \Qhat_{1} \, \rho(t) + \, \rho(t) \, \Qhat_{1} \, \bigg]  \;.
\end{align}
Using \eqref{MatRepLstar} the matrix representing $\Lstar$ reads:
\begin{align}
	(\Lstar) = \left( \begin{array}{cccc}  -k_{21} & 0         & 0          & 0  \\
	                                             0 & -k_{21}/2 & 0          & 0  \\
                                               0 & 0         & - k_{21}/2 & 0  \\
	                                             0 & 0         & 0          & 0  \end{array} \right) \;.
\end{align}
This is clearly not an invertible matrix since the last row has only zeros. However, it has a Moore--Penrose pseudoinverse given by the same matrix with the first three diagonal entries inverted. In operator form we have
\begin{align}
\label{LstarPMInv}
	\Lstar^+ \rho = - \, k^{-1}_{21} \, \big( \hat{Q}_1 \, \rho \, \hat{Q}_1 + 2 \,  \hat{Q}_1 \, \rho \, \hat{Q}_2 + 2 \, \hat{Q}_2 \, \rho \, \hat{Q}_1 \big) \;.
\end{align}
Repeated application of \eqref{LstarPMInv} then gives
\begin{align}
\label{Lstar2rhoi}
	\Lstar^+{}^2 \rhoi = {}& k^{-2}_{21} \big( \hat{Q}_1 \, \rhoi \, \hat{Q}_1 + 4 \,  \hat{Q}_1 \, \rhoi \, \hat{Q}_2 + 4 \, \hat{Q}_2 \, \rhoi \, \hat{Q}_1 \big)\;,  \\
\label{Lstar3rhoi}
	\Lstar^+{}^3 \rhoi = {}& - k^{-3}_{21} \big( \hat{Q}_1 \, \rhoi \, \hat{Q}_1 + 8 \,  \hat{Q}_1 \, \rhoi \, \hat{Q}_2 + 8 \, \hat{Q}_2 \, \rhoi \, \hat{Q}_1 \big) \;.
\end{align}
Taking the average of \eqref{Lstar2rhoi} and \eqref{Lstar3rhoi} against $\ket{\psi_1}$ we get
\begin{align}
	\bra{\psi_1} \Lstar^+{}^2 \rhoi \ket{\psi_1} = \frac{1}{k^2_{21}} \; \rho_{11}  \;, \quad \bra{\psi_1} \Lstar^+{}^3 \rhoi \ket{\psi_1} = -\frac{1}{k^3_{21}} \; \rho_{11}  \;,
\end{align}
where we have defined $\rho_{mn}=\bra{\psi_m}\rhoi\ket{\psi_n}$. The average and variance of $T(\hat{Q}_2,\rhoi)$ according to \eqref{TwoStateE[T]} and \eqref{TwoStateV[T]} are thus 
\begin{gather}
	{\rm E}\big[ \, T(\hat{Q}_2,\rhoi) \, \big] = \frac{1}{k_{21}} \; \rho_{11} \;,  \\
	{\rm V}\big[ T(\hat{Q}_2,\rhoi) \big] = \frac{1}{k^2_{21}} \; ( 2 \rho_{11} - \rho^2_{11} )  \;.
\end{gather}
Note the mean and variance are independent of any initial coherences in the graph so the random walk is classical. This can be expected because incoherent transitions do not couple the site populations to its coherences [see \eqref{ADmatrix} of Sec.~\ref{DmnRho}]. For $\rhoi=\op{\psi_1}{\psi_1}$ we have $\rho_{11}=1$ and
\begin{align}
	{\rm E}\big[ \, T(\hat{Q}_2,\hat{Q}_1) \, \big] = \frac{1}{k_{21}}  \;,  \quad
	{\rm V}\big[ T(\hat{Q}_2,\hat{Q}_1) \big] = \frac{1}{k^2_{21}}  \;.
\end{align}
This says the average amount of time we would have to wait to see the walker reach $\ket{\psi_2}$ starting at $\ket{\psi_1}$ is the inverse of the incoherent transition rate and this has a spread equal to the square of the mean as expected. If on the other hand $\rhoi=\op{\psi_2}{\psi_2}$, then $\rho_{11}=0$, and we obtain 
\begin{align}
\label{AbsRecurrenceTime}
	{\rm E}[T(\hat{Q}_2,\hat{Q}_2)] = {\rm V}\big[ T(\hat{Q}_2,\hat{Q}_2) \big] = 0 \;.
\end{align} 
When the initial and final states coincide, i.e.~$\rhof=\rhoi$, the quantity $T(\rhoi,\rhoi)$ is known as the recurrence time of $\rhoi$ \cite{KMT12}. It is the time taken by a walker starting at $\rhoi$ to return to it. If $\rhoi$ has an infinite recurrence time then it is referred to as a null-recurrent state, and positive recurrent if its recurrence time is finite. An absorbing state $\rhoi$ can thus be seen as a special case of a positive-recurrent state with zero recurrence time since the walker never leaves $\rhoi$. This is precisely what \eqref{AbsRecurrenceTime} says. The variance should also be zero as there should be no spread in the recurrence time of an absorbing state. Although simple, this example illustrates the fact that we actually just require the inverse of the nonzero block in $\Lstar$ that describes the transitions between $\ket{\psi_1}$ and $\ket{\psi_2}$. The reason why the last row (and column) in $\Lstar$ is zero is simply because there are no transitions out of $\ket{\psi_2}$.

\subsubsection{Example 2: A recurrent state for $N=2$}
\label{Recurrent}

We contrast the example above with the graph shown in Fig.~\ref{TwoStateExample} where another incoherent transition from $\ket{\psi_2}$ to $\ket{\psi_1}$ has been added. In this case we modify \eqref{TwoStateLbar1} to 
\begin{align}
\label{TwoStateLbar2}
	\Lstar \, \rho(t) = - \frac{k_{21}}{2} \, \bigg[ \, \Qhat_{1} \, \rho(t) + \, \rho(t) \, \Qhat_{1} \, \bigg]  
	                    + k_{12} \, \bigg[ \Qhat_{12} \, \rho(t) \, \hat{Q}\dg_{12} \,- \frac{1}{2} \, \Qhat_{2} \, \rho(t) -\frac{1}{2} \, \rho(t) \, \Qhat_{2} \, \bigg]\;.
\end{align}
The matrix representation of $\Lstar$ can be shown to be
\begin{align}
\label{Lstar12}
	(\Lstar)  = \left(\begin{array}{cccc}  -k_{21}  &  0                   &  0                   &  k_{12}     \\ 
	                                       0        &  -(k_{21}+k_{12})/2  &  0                   &  0          \\
	                                       0        &  0                   &  -(k_{21}+k_{12})/2  &  0          \\
	                                       0        &  0                   &  0                   &  -k_{12}    \end{array} \right)  \;.
\end{align}
It is clear that $(\Lstar)$ has linearly independent columns. Therefore $\Lstar$ has a standard inverse which in operator form is given by
\begin{align}
\label{Lstar12Inv}
	\Lstar^{-1} \rho = - \big[  k^{-1}_{21} \, \hat{Q}_1 + 2 \,(k_{12}+k_{21})^{-1} \, \hat{Q}_1 \, \rho \, \hat{Q}_2 
	                   + 2 \,(k_{12}+k_{21})^{-1} \, \hat{Q}_2 \, \rho \, \hat{Q}_1 + k^{-1}_{12} \, \hat{Q}_2 \, \rho \, \hat{Q}_2 \big] \;.
\end{align}
As before, repeated application of \eqref{Lstar12Inv} gives
\begin{align}
	\Lstar^{-2} \,\rhoi = {}& - \big[  k^{-2}_{21} \, \hat{Q}_1 + (k_{12} \, k_{21})^{-1} \, \hat{Q}_{12} \, \rhoi \, \hat{Q}_{12}\dg
	                          + 4 \,(k_{12}+k_{21})^{-2} \, \hat{Q}_1 \, \rhoi \, \hat{Q}_2  \nn \\
	                        & + 4 \,(k_{12}+k_{21})^{-2} \, \hat{Q}_2 \, \rhoi \, \hat{Q}_1
	                          + k^{-2}_{12} \, \hat{Q}_2 \, \rhoi \, \hat{Q}_2 \big]   \;,  \\[0.25cm]
	\Lstar^{-3} \,\rhoi = {}& - \big[  k^{-3}_{21} \, \hat{Q}_1 + (k^{-1}_{12} \, k^{-2}_{21} + k^{-2}_{12} \, k^{-1}_{21}) \, \hat{Q}_{12} \, \rhoi \, \hat{Q}_{12}\dg
	                          + 8 \,(k_{12}+k_{21})^{-2} \, \hat{Q}_1 \, \rhoi \, \hat{Q}_2 \nn \\
	                        & + 8 \,(k_{12}+k_{21})^{-2} \, \hat{Q}_2 \, \rhoi \, \hat{Q}_1 
	                          + k^{-3}_{12} \, \hat{Q}_2 \, \rhoi \, \hat{Q}_2 \big]   \;.
\end{align}
Substituting these into \eqref{TwoStateE[T]} and \eqref{TwoStateV[T]} and simplifying we get
\begin{gather}
\label{TwoStateExampleE[T]}
	{\rm E}\big[ \, T(\hat{Q}_2,\rhoi) \, \big] = \frac{1}{k_{21}} \; \rho_{11} + \left( \frac{1}{k_{21}} + \frac{1}{k_{12}} \right) \, \rho_{22}  \;,  \\
\label{TwoStateExampleV[T]}
	{\rm V}\big[ \, T(\hat{Q}_2,\rhoi) \, \big] = \frac{1}{k^2_{12}} \; \big( 2 \rho_{22} - \rho^2_{22} \big)  + \frac{1}{k^2_{21}}   \;.
\end{gather}
Looking at Fig.~\ref{TwoStateExample}, we can see that if $\rhoi=\op{\psi_1}{\psi_1}$ then we would expect the average hitting time to be identical to the case without the transition from $\ket{\psi_2}$ to $\ket{\psi_1}$. This is indeed what we find on setting $\rho_{11}=1$ and $\rho_{22}=0$ in \eqref{TwoStateExampleE[T]} and \eqref{TwoStateExampleV[T]}. If instead we now set $\rhoi=\op{\psi_2}{\psi_2}$ so that $\rho_{11}=0$ and $\rho_{22}=1$, we obtain the mean and variance for the recurrence time of $\ket{\psi_2}$,
\begin{align}
\label{AvgRecTime}
	{\rm E}\big[ \, T(\hat{Q}_2,\hat{Q}_2) \, \big] = \frac{1}{k_{21}} + \frac{1}{k_{12}}  \;, \quad
	{\rm V}\big[ \, T(\hat{Q}_2,\hat{Q}_2) \, \big] = \frac{1}{k^2_{21}} + \frac{1}{k^2_{12}}  \;.
\end{align}
Intuitively we would say the round trip time $T(\hat{Q}_2,\hat{Q}_2)$ should be the sum of the hitting times for each direction:
\begin{align}
\label{T21+T12}
	T(\hat{Q}_2,\hat{Q}_2) = T(\hat{Q}_1,\hat{Q}_2) + T(\hat{Q}_2,\hat{Q}_1)  \;.
\end{align}
Taking the average of \eqref{T21+T12} then reproduces the average in \eqref{AvgRecTime}. Physically we also expect $T(\hat{Q}_1,\hat{Q}_2)$ to be independent of $T(\hat{Q}_2,\hat{Q}_1)$, so taking the variance of \eqref{T21+T12} we get ${\rm V}[T(\hat{Q}_1,\hat{Q}_2)] + {\rm V}[T(\hat{Q}_2,\hat{Q}_1)]$ which is precisely the variance seen in \eqref{AvgRecTime}.

\subsubsection{Example 3: Effects of coherence for $N=4$}

\begin{figure}[t]
\centerline{\includegraphics[width=1\textwidth]{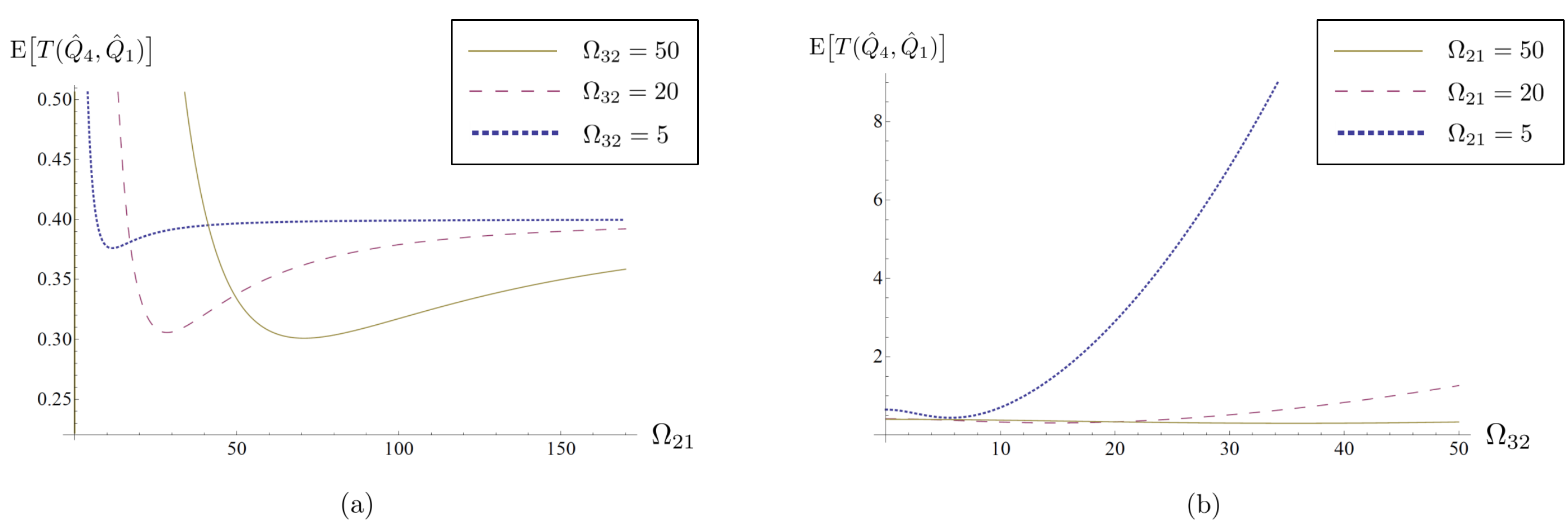}}
\caption{\label{AverageHittingTime} The average hitting time for the graph in Fig.~\ref{HTDwithoutDephasingA} with the parameter values $k_{42}=k_{43}=5$, $\omega_1=1$, $\omega_2=3$, and $\omega_3=5$. (a) We set $\Omega_{32}=5,20,50$ as shown and change only $\Omega_{21}$. (b) $\Omega_{21}=5,20,50$.}
\end{figure}

So far we have only illustrated our formula for the average hitting time for the simplest of examples---graphs with only incoherent transitions and with the minimum number of nodes. Although they are oversimplified, these examples allow us to get a handle on \eqref{E[t]MPInv} by using our intuition from classical random walks. Now that we are somewhat more comfortable with \eqref{E[t]MPInv} we consider a more nontrivial example. We return to the graph discussed in Fig.~\ref{HTDwithoutDephasingA}(a) and examine how different amounts of coherence in the graph can change the average of $T(\hat{Q}_4,\hat{Q}_1)$. We have deliberately chosen the graph parameters so they match those used in Fig.~\ref{HTDwithoutDephasingA}. The parameters which are held constant have their values shown in the caption of Fig.~\ref{AverageHittingTime}.

In Fig.~\ref{AverageHittingTime}(a) we plot the average hitting time as a function of $\Omega_{21}$ for three different values of $\Omega_{32}$ [corresponding to the three values chosen in Fig.~\ref{HTDwithoutDephasingA}(b), (c), and (d)]. It can be seen from Fig.~\ref{AverageHittingTime}(a) that the average hitting times all behave, qualitatively the same as $\Omega_{21}$ is varied for the three values of $\Omega_{32}$ considered: They all  start at infinity at $\Omega_{21}=0$, decay to some minimum, and then increase again to some constant value. From Fig.~\ref{HTDwithoutDephasingA}(a) we see that if $\Omega_{21}=0$ then $\ket{\psi_1}$ becomes an absorbing state and the quantum walker stays at its initial position forever. This is why ${\rm E}[T(\hat{Q}_4,\hat{Q}_1)] \longrightarrow \infty$ as $\Omega_{21} \longrightarrow 0$. Thus as we increase $\Omega_{21}$ from zero, we allow the quantum walker to access the rest of the graph and the average hitting time must drop from infinity. In particular it reaches the minimum when there is non-negligible population in both states $\ket{\psi_2}$ and $\ket{\psi_3}$. Increasing $\Omega_{21}$ further makes the $\ket{\psi_1} \longleftrightarrow \ket{\psi_2}$ transition dominant which leads to a small probability of finding the walker in $\ket{\psi_3}$. Hence when $\Omega_{21} \gg \Omega_{32}$, essentially all first transitions to the final state are from $\ket{\psi_2}$ and the hitting time becomes independent of the actual values of $\Omega_{21}$ and $\Omega_{32}$.

A much less trivial feature is present in Fig.~\ref{AverageHittingTime}(b). This time we fix the value of $\Omega_{21}$ and plot the hitting time as a function of $\Omega_{32}$. Again, all the curves are qualitatively the same and show that $\Omega_{32} \gg \Omega_{21}$ also \emph{increases} the hitting time and can even make it diverge. We emphasise that diverging hitting times are a purely quantum phenomenon as a classical walker always reaches the final state in a finite time (assuming of course that there are no absorbing states). We can shed some light on this quantum effect by solving the eigenvalue problem of the Hamiltonian driving the $\ket{\psi_1} \longleftrightarrow \ket{\psi_2}$ and $\ket{\psi_2} \longleftrightarrow \ket{\psi_3}$ transitions. Since $\omega_1$, $\omega_2$, and $\omega_3$ are of the same order, let us assume they are all equal to zero for simplicity. In this case one can easily show that for $\Omega_{32} \gg \Omega_{21}$ the state $\ket{\psi_1}$ becomes the eigenstate of the Hamiltonian and hence it is a stationary state. In other words, by increasing the rate of coherent transitions $\ket{\psi_2} \longleftrightarrow \ket{\psi_3}$ we are trapping the population in the initial state $\ket{\psi_1}$, and thereby increasing the hitting time. This new mechanism leading to diverging hitting times should be contrasted with two other mechanisms known previously in the context of measured quantum walks \cite{VKB08}: (i) When the measurement frequency tends to zero (since the walk is essentially not measured at all the hitting time tends to infinity); (ii) when the measurement frequency tends to infinity (quantum Zeno effect freezes the population in the initial state).

\section{Comparison with a discrete-time measured walk}  
\label{SEC_COMPARISON}

\subsection{Generalised Krovi--Brun definition}

There are a few noteworthy differences between what we have considered thus far and what is usually considered in the standard theory of quantum walks: First, much of quantum-walk theory considers only closed systems i.e.~systems undergoing unitary evolution whereas the theory presented here allows for non-unitary evolution (although open quantum walks are starting to attract some attention recently as already mentioned in Sec.~\ref{Intro}). Second, time is often discrete. In this case the hitting time is simply measured in the number of steps that the system takes to reach the final state. The hitting probability is then defined analogously to \eqref{GeneralHTD} to be 
\begin{align}
\label{f(n)Defn}
	f(n;\rhof,\rhoi) \equiv {\rm Pr}\big[ \rho(t_n) = \rhof, \rho(t_m) \ne \rhof  \;\forall \; m \le n-1 \, | \, \rho(0) = \rhoi  \big] \;.
\end{align}
Lastly, as we said already for quantum walks with only unitary dynamics, the notion of a quantum walker actually being in a particular state is not well defined. One way around this issue is to introduce a measurement of the final-state population at the end of every step. Given that our hitting distribution is founded on measurement theory, it is most natural to compare our calculation with this version of the hitting distribution from the quantum-walk literature. In particular, we use the definition stated in Krovi and Brun's paper \cite{KB06} but replace, in their definition of the hitting distribution, the unitary time-evolution operator by a non-unitary map $\Ks(t)$. Taking $\rhof=\hat{Q}_N$ as before, the expression for $f(n;\rhof,\rhoi)$ from Ref.~\cite{KB06} gives 
\begin{align}
\label{f(n)}
	f(n;\hat{Q}_N,\rhoi) = {\rm Tr}\big\{ {\cal Q}_N \, \Ks(\dt) \big[ {\cal P}_N \, \Ks(\dt) \big]^{n-1} \rhoi \big\}
\end{align}
where the quantum-walk is defined by ${\cal K}(\dt)=\exp(\Ls\,\dt)$. The superoperators ${\cal Q}_N$ and ${\cal P}_N$ are defined by 
\begin{align}
\label{QandP}
	{\cal Q}_N \, \rho = \hat{Q}_N \, \rho \,\hat{Q}_N \;, \quad {\cal P}_N \, \rho = \hat{P}_N \, \rho \, \hat{P}_N  \;, 
\end{align}
and
\begin{align}
\label{PhatN}
	\hat{P}_N = \hat{1} - \hat{Q}_N \;.
\end{align}
The effect of ${\cal Q}_N$ is thus to project the system into $\ket{\psi_N}$ while ${\cal P}_N$ projects the system into the subspace  
\begin{align}
	\bar{\mathbb{H}}_N \equiv {\rm Span}\big(\bar{\mathbb{S}}_N \big) \;, 
\end{align}
where $\bar{\mathbb{S}}_N \equiv \mathbb{S}_N - \{ \ket{\psi_N} \}$. Their effects on an arbitrary state $\rho$ can most easily be seen when \eqref{QandP} are in their matrix representations 
\begin{align}
\label{QandPMatrixForm}
	{\cal Q}_N \, \rho = \left(\begin{array}{ccccc} 0      & 0      & \cdots & 0      & 0         \\ 
	                                                  0      & 0      & \cdots & 0      & 0         \\
	                                                  \vdots & \vdots &        & \vdots & \vdots    \\
	                                                  0      & 0      & \cdots & 0      & 0         \\
	                                                  0      & 0      & \cdots & 0      & \rho_{NN}  \end{array} \right)  \;,  \quad
	{\cal P}_N \, \rho = \left(\begin{array}{ccccc} \rho_{11}    & \rho_{12}    & \cdots & \rho_{1,N-1}   & 0       \\ 
	                                                  \rho_{21}    & \rho_{22}    & \cdots & \rho_{2,N-1}   & 0       \\
	                                                  \vdots       & \vdots       &        &  \vdots        & \vdots  \\
	                                                  \rho_{N-1,1} & \rho_{N-1,2} & \cdots & \rho_{N-1,N-1} & 0       \\
	                                                  0            & 0            & \cdots & 0              & 0  \end{array} \right)  \;,
\end{align}
where we have made use of \eqref{RhoMatrixElements} for the matrix elements of $\rho$ (sometimes with a comma in the subscript when the indices of $\rho$ are otherwise difficult to differentiate). Note that we have written the time step in \eqref{f(n)} as a small but finite number $\dt$ to remind ourselves that \eqref{f(n)} was derived in discrete time. Therefore the comparison of $h(t;\rhof,\rhoi)$ to $f(n;\rhof,\rhoi)$ has to be made in the limit of $\dt \longrightarrow 0$. As a matter of fact, we will show next that the two distributions converge in the continuous-time limit, expressed as
\begin{align}
	\lim_{\dt \to 0} f(n;\hat{Q}_N,\rhoi) = h(t_{n-1}) \, \dt  \;.
\end{align}

\subsection{Convergence of the generalised Krovi--Brun and quantum-jump distributions}

To show the equivalence between the two hitting distributions in the $\dt \longrightarrow 0$ limit it helps to separate the full generator for the Markov chain into jump and no-jump terms: 
\begin{align}
	\Ls = \Lstar + \Jstar  \;,
\end{align}
where $\Lstar$ is given by \eqref{SimpleFormLbarWithDeph} and we have also defined the shorthand 
\begin{align}
	\Jstar \equiv {\cal J}(\bm{\eta}_\star)  \;.
\end{align}
Recall that we defined ${\cal J}(\bm{\eta}_\star)$ in \eqref{Jstar}. We will assume that initially there is no population in the final state,
\begin{align}
\label{InitialCondition1}
	\bra{\psi_N} \rhoi \ket{\psi_N} = 0  \;.
\end{align}
This condition then implies that $\ket{\psi_N}$ cannot share coherences with the remaining states in $\bar{\mathbb{S}}_N$. The initial state is thus confined to be in $\bar{\mathbb{H}}_N$, which can be formally expressed as
\begin{align}
\label{InitialCondition2}
	{\cal P}_N \, \rhoi = \rhoi  \;.
\end{align}
The idea of writing $\Ls$ as the sum of $\Lstar$ and $\Jstar$ is that the no-jump and jump superoperators have the following useful properties which can be used to show the consistency between the discrete-time and continuous-time hitting distributions. The first thing to note is that the no-jump evolution of an $\rhoi$ defined by \eqref{InitialCondition2} is confined to $\bar{\mathbb{H}}_N$. This automatically means that any future evolution of $\rhoi$ under $\Lstar$ has zero projection onto $\ket{\psi_N}$. We can express these observations formally as
\begin{align}
\label{LstarIdentity}
	{\cal Q}_N \, \Lstar = 0 \;,  \quad  {\cal P}_N \, \Lstar = \Lstar \, {\cal P}_N = \Lstar \;.
\end{align}
Remember that these are superoperator identities assuming they act on states satisfying \eqref{InitialCondition2}. Jumps, on the other hand, take the system state out of its confinement in $\bar{\mathbb{H}}_N$ and into $\ket{\psi_N}$. This means that
\begin{align}
\label{JstarIdentity}
	{\cal Q}_N \, \Jstar = \Jstar  \;,  \quad  {\cal P}_N \, \Jstar = 0 \;.
\end{align}
We then have, in the limit of $\dt \longrightarrow 0$,
\begin{align}
	{\cal P}_N \, \Ks(\dt) = {}& {\cal P}_N \, e^{\Ls \dt}  \\
                         = {}& {\cal P}_N \big( \mathbbm{1} + \Lstar \, \dt + \Jstar \, \dt \big) \\
                         = {}& \big( \mathbbm{1} + \Lstar \, \dt \big) \, {\cal P}_N  \;.
\end{align}
On rewriting the expansion in $\dt$ back in exponential form we arrive at
\begin{align}                         
	{\cal P}_N \, \Ks(\dt) \, \rhoi = e^{\Lstar \, \dt} \, \rhoi  \;.
\end{align}
This then permits us to write 
\begin{align}
	\bar{\rho}(t_{n-1}) =\big[ {\cal P}_N \, \Ks(\dt) \big]^{n-1} \rhoi = e^{\Lstar (n-1) \dt} \rhoi  \;.
\end{align}
This is again a state which satisfies \eqref{InitialCondition2} [with $\rhoi$ replaced by $\bar{\rho}(t_{n-1})$] so the superoperator identities \eqref{LstarIdentity} and \eqref{JstarIdentity} still apply and we have
\begin{align}
	{\cal Q}_N \, \Ks(\dt) \, \big[ {\cal P}_N \Ks(\dt) \big]^{n-1} \rhoi = {}& {\cal Q}_N \, \Ks(\dt) \, \bar{\rho}(t_{n-1})   \\
                                                                        = {}& {\cal Q}_N \, \big( \mathbbm{1} + \Lstar \, \dt + \Jstar \, \dt \big) \, \bar{\rho}(t_{n-1})  \\
                                                                        = {}& \Jstar \, \bar{\rho}(t_{n-1}) \, \dt  \;,
\end{align}
where we have also noted that ${\cal Q}_N \,\bar{\rho}(t_{n-1})=0$. Taking the trace then gives
\begin{align}         
\label{DiscreteTimeJumpProb}                
	f(n)= {\rm Tr}\big[ \Jstar \, \bar{\rho}(t_{n-1}) \big] \, \dt  \;,
\end{align}
which we recognise as the jump probability \eqref{JumpProb} with an unnormalised state, but we showed in \eqref{ConditionedHTD} and \eqref{NoJumpProbability} that this is precisely the hitting distribution. The only difference is that time is discretised. Substituting the definition of $\Jstar$ from \eqref{Jstar} into \eqref{DiscreteTimeJumpProb} we get
\begin{align}                         
	f(n) = {}& {\rm Tr}\Bigg[ \sum_{m=1}^{N-1}  \, k_{Nm} \, \hat{Q}_{Nm} \, \bar{\rho}(t_{n-1}) \; \hat{Q}\dg_{Nm} \Bigg] \, \dt  \\
       = {}& \sum_{m=1}^{N-1}  \, k_{Nm} \, \bra{\psi_m} \, e^{\Lstar (n-1) \dt} \rhoi \; \ket{\psi_m} \, \dt = h(t_{n-1}) \, \dt  \;.
\end{align}
Note that $f(n)$ corresponds to the quantum-jump distribution evaluated at time $t_{n-1}$. This makes sense since there are only $n$ time steps so what we want is the probability that a jump occurs in the interval from $t_{n-1}$ to $t_n$ and that none occurred from $t_0$ to $t_{n-1}$---which is precisely $h(t_{n-1})\,\dt$.

\section{Including coherent transitions to the final state}
\label{N+1Model}

\subsection{The $N+1$ model}
\label{NewModel}

The theory above does not allow for coherent transitions to $\ket{\psi_N}$ in the graph. The quantum-jump approach works by detecting transitions to the final state but only incoherent transitions can be detected. That is to say, if there are coherent transitions from $\bar{\mathbb{H}}_N$ to $\ket{\psi_N}$, then the hitting statistics calculated from the quantum-jump approach will not reflect the true statistics, which contains a contribution from the coherent transitions to the final state. On the other hand the Krovi--Brun formula \eqref{f(n)} works by detecting the on-site populations and therefore applies to arbitrarily complicated graphs. We will now remedy this problem so that the quantum-jump approach can be applied to arbitrarily complicated graphs as well.

Let us introduce one extra state, a fictitious state, to the quantum walk. We will denote this extra state as $\ket{\psi_{N+1}}$. By introducing an incoherent transition from $\psiN$ to $\psiNp$ we can expect to recover the original hitting time (i.e.~the time to reach $\psiN$ for the first time) by making the transition rate $k_{N+1,N}$ large in the extended graph. Since the rate $k_{N+1,N}$ is an important quantity in this section, and is also a bit cumbersome to write repeatedly, we shall set
\begin{align}
	v \equiv k_{N+1,N}  \;.
\end{align}
It is then intuitive to see that
\begin{align}
	\lim_{v \to \infty} T\big( \hat{Q}_{N+1},\rhoi \big) = T\big( \hat{Q}_N,\rhoi \big)  \;,
\end{align}
since in the limit of large $v$, the time taken to go from $\psiN$ to $\psiNp$ becomes negligible. In fact, in the limit of large $v$, we can also expect $T(\hat{Q}_{N+1},\rhoi)$ and $T(\hat{Q}_N,\rhoi)$ to be identically distributed. This is because as we make $v$ larger and larger, we are also making the probability of reaching $\psiNp$ from $\psiN$ tend to one. When this probability is one, the quantum walker is as likely to be in $\psiNp$ as $\psiN$, or more formally,
\begin{align}
\label{EquivalenceOfHTDs}
	\lim_{v \to \infty} h(t;\hat{Q}_{N+1},\rhoi) \, dt = \lim_{\dt \to 0} f(n;\hat{Q}_N,\rhoi) \;.
\end{align}
Note the hitting distribution to reach $\psiN$ is written using the Krovi--Brun formula since the existence of coherent transitions to $\psiN$ limits the use of the quantum-jump method. We refer to the Krovi--Brun formula $f(n;\hat{Q}_N,\rhoi)$ simply as the $N$ model and the quantum-jump formula $h(t;\hat{Q}_{N+1},\rhoi) \, dt$ in the limit of large $v$ as the $N+1$ model. We prove \eqref{EquivalenceOfHTDs} below.

\subsection{Convergence of the $N$ and $N+1$ models}

By adding one extra state and an incoherent connection from $\psiN$ to $\psiNp$ we may decompose the dynamics of the $N+1$ graph into jump and no-jump terms as we have been doing,
\begin{align}
\label{OldNotation}
	\Ls = \Lstar + \Jstar  \;.
\end{align}
Note that \eqref{OldNotation} now acts on density operators in an $N+1$-dimensional space where $\Lstar$ describes the evolution of the quantum walk constrained to the original $N$-dimensional subspace. The superoperator $\Jstar$ now describes a jump from $\psiN$ to $\psiNp$, defined by
\begin{align}
	\Jstar \, \rho = v \: \hat{Q}_{N+1,N} \, \rho \, \hat{Q}\dg_{N+1,N}  \;.
\end{align}
In order to make the connection with the $N$ model, it helps to introduce
\begin{align}
\label{NewNotation}
	\Ls = \Ls_{N+1} \;, \quad  \Lstar = \Ls_N  \;.  
\end{align}
The proof of \eqref{EquivalenceOfHTDs} can be made easier if we discretise time into steps of size $\dt$ and express $h(t)\,dt$ in the limit of $\dt \longrightarrow 0$. In this case an arbitrary $t$ is equivalent to $t_n\equiv n\,\dt$ for an arbitrary $n$. The hitting distribution calculated using quantum jumps is thus
\begin{align}
	h(t_n,\hat{Q}_{N+1},\rhoi) \, \dt = {}& {\rm Tr}\Big\{ \Jstar \big[ e^{\Lstar \dt} \big]^n \rhoi \Big\} \, \dt  \\
                                    = {}& {\rm Tr}\Big\{ \Jstar \, e^{\Lstar \dt} \big[ e^{\Lstar \dt} \big]^{n-1} \rhoi \Big\} \, \dt  \\  
                                    = {}& {\rm Tr}\Big\{ \Jstar \, e^{\Ls_N \,\dt} \big[ {\cal P}_{N+1} \, e^{\Ls_{N+1} \, \dt} \, \big]^{n-1} \rhoi \Big\} \, \dt  \;, 
\end{align}
where ${\cal P}_{N+1}$ is defined in the same way as \eqref{QandP}, \eqref{PhatN}, and \eqref{QandPMatrixForm} with $N \longrightarrow N+1$ [recall also that we have defined $\hat{Q}_n=\op{\psi_n}{\psi_n}$]. The last line simply makes use of the proof in Sec.~\ref{SEC_COMPARISON} to express the conditional evolution in terms of ${\cal P}_{N+1}$ and $\Ls_{N+1}$. As we will be taking the limit of large $v$ it makes sense to plug in the definition of $\Jstar$ to make $v$ explicit in the hitting distribution:  
\begin{align}
\label{h(tn)}  
	h(t_n,\hat{Q}_{N+1},\rhoi) \, \dt = v \, \dt \; {\rm Tr}\Big\{ {\cal Q}_N \, e^{\Ls_N \,\dt} \big[ {\cal P}_{N+1} \, e^{\Ls_{N+1} \, \dt} \, \big]^{n-1} \rhoi \Big\} \;.
\end{align}
There are in fact two places where $v$ appears in \eqref{h(tn)}: One in the product $v \, \dt$, and the other implicitly in $\exp(\Ls_{N+1}\,\dt)$. The first limit is simple to see. All we have to do is note that \eqref{h(tn)} is already in the $\dt \longrightarrow 0$ limit and that $v \longrightarrow \infty$ must be consistent with the interpretation of $v \, \dt$ as the probability to jump from $\psiN$ to $\psiNp$ in $\dt$. This means that $v \, \dt \longrightarrow 1$ as we make $v$ ever so large but $\dt$ ever so small. This immediately gives
\begin{align}
\label{vLimit}
	\lim_{v \to \infty} h(t_n,\hat{Q}_{N+1},\rhoi) \, \dt
	= {\rm Tr}\Big\{ {\cal Q}_N\, e^{\Ls_N \,\dt} \Big[ {\cal P}_{N+1} \, \lim_{v \to \infty} e^{\Ls_{N+1} \, \dt} \, \Big]^{n-1} \rhoi \Big\}  \;.
\end{align}
The only nontrivial part now resides in what happens to the evolution of the $N+1$ graph for large $v$: The larger we make $v$, the more difficult it is for the population of $\psiN$ to build up. Then, as $v$ tends to infinity there is simply no appreciable population in $\psiN$ for all time. Since $\bra{\psi_N}\rho(t)\ket{\psi_N}=0$ implies no coherences can be shared between $\psiN$ and any other state, we can write, in the $v \longrightarrow \infty$ limit,
\begin{align}
\label{AdiabaticApprox}
	e^{\Ls_{N+1} \, \dt} = {}& {\cal P}_{N} \, e^{\Ls_{N+1} \, \dt} \;. 
\end{align}
This simply says that the $N^{\rm th}$ row and $N^{\rm th}$ column of the density operator at all times is zero when $v \longrightarrow \infty$. Now we note that the only sensible initial state is one satisfying
\begin{align}
	{\cal P}_{N+1} \, \rhoi = \rhoi  \;,
\end{align}
since the original hitting problem is defined only for the $N$-state graph. From this initial condition it is easy to see that  
\begin{align}
\label{ConditionsOnPN+1}
	{\cal P}_{N+1} \, \Ls_N = \Ls_N  \, {\cal P}_{N+1} = \Ls_N  \;,   \quad  {\cal P}_{N+1} \, \Jstar = 0   \;.
\end{align}
The properties in \eqref{ConditionsOnPN+1} are simple to see since the action of ${\cal P}_{N+1}$ just says that we find the system to be in a state in the subspace spanned by $\mathbb{S}_N$. The first condition in \eqref{ConditionsOnPN+1} then follows since $\Ls_N$ only evolves the system within the subspace spanned by $\mathbb{S}_N$. The second condition in \eqref{ConditionsOnPN+1} is true because $\Jstar$ puts the quantum walker in the state $\ket{\psi_{N+1}}$ so that one finds the population of every other state to be zero. These are the same conditions as those appearing in \eqref{LstarIdentity} and \eqref{JstarIdentity}, just extended to $N+1$ states. We can use \eqref{ConditionsOnPN+1} to simplify the limit in \eqref{vLimit} by noting that 
\begin{align}
\label{Commutability}
	{\cal P}_{N+1} \, {\cal P}_N = {\cal P}_N \, {\cal P}_{N+1}  \;.
\end{align}
We therefore have, upon using \eqref{ConditionsOnPN+1} and \eqref{Commutability},
\begin{align}
	{\cal P}_{N+1} \, {\cal P}_{N} \, e^{\Ls_{N+1} \, \dt} = {}& {\cal P}_{N} \, {\cal P}_{N+1} \big( \mathbbm{1} + \Ls_N \, \dt + \Jstar \, \dt \big)  \\
                                                         = {}& {\cal P}_{N} \, e^{\Ls_{N} \, \dt}\, {\cal P}_{N+1} \;.                                                        
\end{align}
Since ${\cal P}_{N+1}$ can be commuted through ${\cal P}_N$ and $\Ls_N$, it is simple to see that 
\begin{align}
	\big[ {\cal P}_{N+1} \, {\cal P}_{N} \, e^{\Ls_{N+1} \, \dt} \big]^{n-1} = \big[ {\cal P}_{N} \, e^{\Ls_{N} \, \dt} \, \big]^{n-1} \, {\cal P}_{N+1} \;.                                                        
\end{align}
Substituting this back into \eqref{vLimit} then gives, for $v \longrightarrow \infty$ and $\dt \longrightarrow 0$,
\begin{align}
	h(t_n,\hat{Q}_{N+1},\rhoi) \, \dt
	= {\rm Tr}\Big\{ {\cal Q}_N\, e^{\Ls_N \,\dt} \big[ {\cal P}_N \, e^{\Ls_{N} \, \dt} \, \big]^{n-1} \rhoi \Big\}  \;.
\end{align}
As this equation already assumes $\dt \longrightarrow 0$, this is precisely the right-hand side of \eqref{EquivalenceOfHTDs}. This therefore shows the equivalence between the $N$ and $N+1$ models.
\begin{figure}[t]
\centerline{\includegraphics[width=0.9\textwidth]{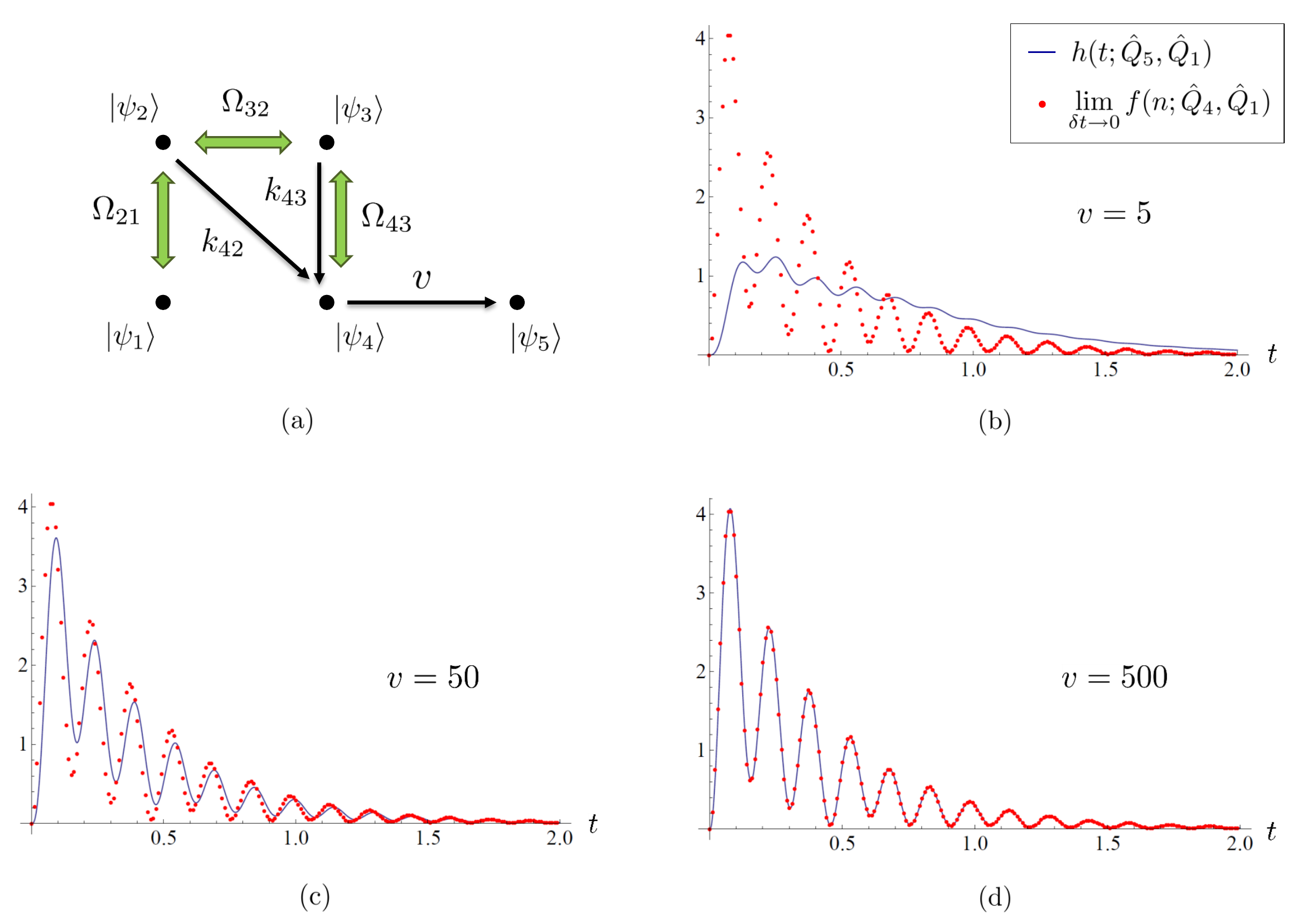}}
\caption{\label{NandN+1} Hitting-time distributions for the $N+1$ and $N$ models for the $N=4$ graph shown in (a) ($\ket{\psi_5}$ being a fictitious state). The parameter values are $\Omega_{43}=\Omega_{32}=k_{42}=k_{43}=5$, $\Omega_{21}=50$, $\omega_1=1$, $\omega_2=3$, and $\omega_3=5$. The convergence between the $N+1$ and $N$ models is clearly seen as $v$ is increased from $v=5$ in (b) to $v=500$ in (d).}
\end{figure}

We now illustrate \eqref{EquivalenceOfHTDs} for the graph in Fig.~\ref{NandN+1}(a) where we are interested in the hitting time for vertex $\ket{\psi_4}$ starting at $\ket{\psi_1}$. All the graph parameters except for $v$ are fixed and have values stated in the figure caption. In this case we consider the hitting-time distribution for $\ket{\psi_5}$ under the quantum-jump approach and compare it to the generalised Krovi--Brun distribution taking the final state to be $\ket{\psi_4}$. The hitting distribution calculated from the quantum-jump approach is shown as the (blue) solid line while the generalised Krovi--Brun approach is shown as (red) dots. The value of $v$ is gradually increased as we pass from Fig.~\ref{NandN+1}(b) to (d), and as can be seen, the hitting distribution from the quantum-jump approach (solid curve) changes until it eventually coincides exactly with the generalised Krovi--Brun hitting distribution (dots).

\section{Discussion}
\label{Conclusion}

The most essential result of this paper is the explicit expression for the statistics of hitting times for a model of continuous-time open quantum walks given in \eqref{M(x)Final}. This is a finite sum and in principle can be entered into a computer to extract meaningful quantities about the quantum walk such as the average hitting time and its variance. Below we summarise our results in conjunction with other findings obtained in this paper. We also explain the relationship between our work and some other literature which we have not mentioned so far but are nevertheless worthy of a discussion.

\subsection{Summary}

We have shown here how the quantum-jump method lends itself naturally to the hitting problem in continuous open quantum walks. In this language the hitting time for the walker to reach a prescribed state $\rhof$ from some fixed $\rhoi$ in an $N$-state graph is defined by the time of the first jump. Based on this, the hitting distribution is just the distribution of times for the first jump, given in \eqref{FinalHTD}. Simple examples of this result were considered. We then derived expressions for the statistical moments of the hitting time in terms of the graph dynamics. The results are given by \eqref{E[t]MPInv}, \eqref{V[T]Final}, and \eqref{M(x)Final}. Again, some simple examples were used to illustrate the final results. In the process we showed analytically how the quantum expression simplifies to the expected classical result when the graph has only incoherent transitions. We then generalised the hitting distribution for the discrete-time unitary quantum walk in Ref.~\cite{KB06} to allow for non-unitary evolution and showed that this is equivalent to the quantum-jump distribution if the time step of the discrete walk approached zero. A caveat of the quantum-jump method is that it assumes the edges connected to $\rhof$ are incoherent. The hitting time becomes much less straightforward to define when coherent transitions to $\rhof$ are present. Thus open quantum walks have an intrinsic advantage over unitary walks in that there is a set of graphs for which hitting times can be unambiguously defined (i.e.~graphs where $\rhof$ only shares incoherent edges with other nodes). We then proposed a solution to this problem by adding one fictitious state to the graph giving us a model with $N+1$ states. We then proved that our $N+1$ model predicts the same hitting statistics as the $N$ model according to the generalised definition of Ref.~\cite{KB06}. This is illustrated for a simple graph where the convergence of the $N+1$ model to the $N$ model can be seen ``explicitly'' (Fig.~\ref{NandN+1}). It is interesting to point out that the $N$ and $N+1$ models differ qualitatively in that the former decides if the final state has been hit or not by detecting the walker's presence \emph{at} the final state, whereas the $N+1$ model works by detecting if the walker is \emph{in transit} to the final state.

\subsection{Relation to other work and possible future explorations}
\label{FurtherDiscussions}

Our appeal to photon counting as a way of thinking about quantum jumps for the hitting problem raises another interesting question related to the use of measurements in quantum walks. This is related to our allusion to other unravellings made earlier in Sec.~\ref{QuantumJumps}. In general, an unravelling is a specific decomposition of the master equation into stochastic trajectories such that the ensemble average of them recovers the dynamics of the master equation. Two classes of unravellings have special significance in quantum optics---the jump unravellings used in our paper, and the so-called diffusive unravellings---because they turn out to have concrete interpretations in terms of quantum optical measurements (as already seen with the jump unravellings in our work) \cite{WM10}. Canonical examples of quantum-optical measurements that give rise to (or essentially realise) diffusive unravellings are the homodyne and heterodyne detection schemes \cite{WD01,CW11}. Since we are defining the graph of a continuous-time open quantum walk by a master equation, and we know that a master equation can be unravelled in different ways, one might wonder if another type of unravelling can be used to study a hitting problem. Given that diffusive unravellings form the second major class of unravellings in quantum optics we might then ask if they can be used for the hitting problem in this paper. The short answer is that there is not much motivation to do so because they are unhelpful for solving hitting problems with discrete-state quantum walks. To understand this we recall that in our quantum-jump approach we imagined the incoherent transitions in the graph gave off photons which were measured by photodetectors. Using a diffusive unravelling is equivalent to superimposing the stream of photons coming from an incoherent transition with a local-oscillator field (a laser in a coherent state) on a beam splitter before it is detected by the photodetector. In this case the photodetector actually measures a quadrature of the stream of photons determined by the phase of the local oscillator. A diffusive unravelling thus gives us information about a continuous variable that does not directly reveal to us if the walker has reached a prescribed state or not. To get around this one can then try to introduce the fidelity between $\rhof$ (here taken to be a parameter) and the walker's state to define when $\rhof$ is reached or not. As one can see, this already introduces an extra step into the analysis, not to mention that the fidelity for a fixed $\rhof$ is a nonlinear function of the density operator. The jump unravelling on the other hand does give us direct information about whether the walker has reached $\rhof$, and is much more natural for our quantum-walk problem. If one wants to think about applying diffusive unravellings one could start with a completely different quantum-walk model altogether where diffusive unravellings would appear to be natural. A suitable quantum-walk model would be one in which the state space is continuous. To this end, it might be interesting to use quantum Brownian motion as a model. A diffusive unravelling of this can then be expected to reveal the walker's position directly.

There is in fact an example of a hitting-time calculation for a diffusively unravelled quantum master equation in the context of rapid purification \cite{Jac03,Jac04,CJ06}. The basic idea is to purify qubits as quickly as possible by using a continuous diffusive measurement and feedback. There are two approaches to rapid purification---one can either maximize the average purity of a qubit in a fixed time, or minimize the average time for a qubit to reach a fixed purity \cite{WR06}. It is the latter approach that is a hitting problem where the hitting time is simply the time to reach a predefined purity. In this case, the average time to reach a predefined purity (i.e.~the average hitting time) is calculated by first converting the qubit dynamics under a diffusive measurement into a set of stochastic Bloch equations, and then from these, the equivalent classical Fokker--Planck equation. Once the classical Fokker--Planck equation is known, standard techniques for calculating the average time to reach a given purity can be applied \cite{Gar04} (where the hitting time is known as the first-passage time). Aside from the difference in the unravelling that is employed, there is another difference between the calculation of the average first-passage time in the rapid purification work and our results here: Our average hitting-time calculation (and in fact all higher statistics) uses a quantum method---quantum jumps---and hence our results are expressed explicitly in terms of the quantum evolution equation of the random process. The rapid-purification work on the other hand re-expresses the evolution of the random process in terms of the equivalent classical equation first (the Fokker--Planck equation), from which the average hitting/first-passage time can then be obtained using a method invented for classical diffusive processes. In light of this, a noteworthy point of our paper is that the nature of our problem, and the ensuing method used, allows us to bypass the extra step of making a quantum-classical correspondence as in rapid purification before the hitting problem can be solved.

We have seen that in the language of photon counting, the hitting time is the analogue of the time of the first count. In fact, our calculation of the hitting-time distribution follows much the same strategy as what a quantum optician would do to derive the distribution of waiting times (the time interval between successive photon counts). Is the hitting-time distribution in our quantum-walk problem therefore equivalent to the waiting-time distribution of photon counting? The answer is no, and to see this we just have to refer back to Fig.~\ref{JumpsOnGraph}. If we look at the record of jumps shown on the right in Fig.~\ref{JumpsOnGraph}, the waiting time is the interval between successive jumps. However, from the graph on the left we can see that this is the amount of time the walker spends being away from the final state before returning to it again. The time taken for a random walker to start in a given node and return to it turns out to be an interesting quantity as well. It is known as the recurrence time (recall Examples 1 and 2 in Sec.~\ref{Examples}). The recurrence time in a hitting problem is thus the quantity that one should regard as analogous to the waiting time of quantum optics and it is not difficult to see that it is different to the hitting time. It is in fact simple to construct a graph for which the hitting time is very different to the recurrence time. We summarise this in Table~\ref{Analogies}. 
\begin{table}
\begin{center}
	\begin{tabular}{|c|c|}
  \hline 
	Hitting problem    &  Photon counting     \\    
  \hline 
  Hitting time       &  Time of first count     \\ 
  Recurrence time    &  Waiting time      \\
	\hline
	\end{tabular}
\end{center}
\caption{\label{Analogies} Analogies between different quantities in a hitting problem of quantum walks and photon counting in quantum optics.}
\end{table}

It is then interesting to ask whether there is additional knowledge from photodetection statistics that one might be able to borrow for quantum walks. One potential application is the use of the waiting-time distribution for estimating an unknown parameter in a quantum emitter that one is interested in characterising \cite{KM15,KM14}. The simplest model being a driven two-level atom whose Rabi frequency is not known to an experimenter. Kiilerich and M{\o}lmer have shown that the waiting-time distribution of the photodetection record obtained from monitoring the atomic fluorescence is useful for estimating the Rabi frequency \cite{KM14}. This idea may have applications to quantum walks where one has only a partially characterised graph---e.g.~the frequency of one of the coherent transitions in the graph is not known. It might then be possible to identify a recurrent state whose distribution of recurrence times can be used for identifying the unknown transition frequency. Clearly, one will need the multichannel extension of this idea described in Ref.~\cite{KM15}. A successful application of this theory would then extend the quantum-walk--quantum-optics analogy further.

\section{Acknowledgements}

We would like to thank an anonymous referee who stimulated us into thinking about continuous-time quantum walks. This paper is an outgrowth of the exchange we had with the referee. We thank also many others who we have discussed this project with in the past: Itai Arad, Marcin Karczewski, Dagomir Kaszlikowski, Pawe{\l} Kurzy\'{n}ski, Zakarya Lasmar, Gerard Milburn, Ranjith Nair, Changsuk Noh, and Miklos Santha. This research is supported by the MOE grant number RG 127/14, and the National Research Foundation, Prime Minister's Office, Singapore under its Competitive Research Programme (CRP Award No. NRF-CRP-14-2014-02).

\appendix

\section{Independent derivation of the average hitting time}
\label{AppA}

The goal here is to provide a derivation of the average of hitting times that is independent of the moment-generating function used in Sec.~\ref{nthMomentProof}. This will add further confidence in our result. The ensemble average of $T(\rhof,\rhoi)$ with $\rhof=\hat{Q}_N$ is defined by  
\begin{align}
\label{E[t]}
	{\rm E}\;\!\big[\;\!T(\hat{Q}_N,\rhoi)\;\!\big] = \int^\infty_0 dt \; t \: h(t;\hat{Q}_N,\rhoi)  \;,
\end{align}
We will again suppress the dependence of $T$ and $h$ on the initial and final states in the following. We integrate \eqref{E[t]} by parts directly, in which case it will be convenient to define the indefinite integral
\begin{align}
\label{H(t)}
	H(t) = \int dt \; h(t) \;.
\end{align}
We may then write
\begin{align}
	{\rm E}\;\![\;\!T\;\!] = \Big[ \, t \, H(t) \, \Big|^{\infty}_0 - \int^\infty_0 dt \; H(t)  \;.
\end{align}
The first term is given by
\begin{align}
\label{LimtH(t)}
	\Big[ \, t \, H(t) \, \Big|^{\infty}_0 = \lim_{t \to \infty} t  \, H(t) - \lim_{t \to 0} t \, H(t) \;.
\end{align}
We can get some insight into this function by actually doing the indefinite integral in \eqref{H(t)}. The function $H(t)$ is 
\begin{align}
	H(t) = {}& \int dt \, \sum_{m=1}^{N-1} \, k_{Nm} \, \bra{\psi_m} \, e^{\Lstar t} \rhoi \ket{\psi_m}   \\
       = {}& \sum_{m=1}^{N-1} \, k_{Nm} \, \bra{\psi_m} \, \left( \int dt \; e^{\Lstar t} \right) \rhoi \;\! \ket{\psi_m}  \\
\label{H(L)}              
       = {}& \sum_{m=1}^{N-1} \, k_{Nm} \, \bra{\psi_m} \, \Lstar^{-1} \, e^{\Lstar t} \rhoi \;\! \ket{\psi_m}  \;.
\end{align}
Assuming that $\Lstar$ is invertible we can see that
\begin{align}
	\lim_{t\to 0}	H(t) = {}& \sum_{m=1}^{N-1} \, k_{Nm} \, \bra{\psi_m} \, \Lstar^{-1} \lim_{t \to 0} e^{\Lstar t} \rhoi \ket{\psi_m}  \\
                     = {}& \sum_{m=1}^{N-1} \, k_{Nm} \, \, \bra{\psi_m} \, \Lstar^{-1} \, \rhoi \ket{\psi_m}  \;,
\end{align}
which in general has a value between zero and some finite number. Therefore we obtain
\begin{align}
	\lim_{t \to 0} t \, H(t) = 0  \;.
\end{align}

As for the first term in \eqref{LimtH(t)}, we need to see how $H(t)$ behaves as $t \longrightarrow \infty\,$. This is given by
\begin{align}
	\lim_{t \to \infty} H(t) = {}& \sum_{m=1}^{N-1} \, k_{Nm} \, \bra{\psi_m} \, \Lstar^{-1} \lim_{t \to \infty} e^{\Lstar t} \rhoi \, \ket{\psi_m}  = 0  \;,
\end{align}
where we have used \eqref{ZeroSteadyState}, stated here again for ease of reference,
\begin{align}
\label{RhossDefn}
	 \lim_{t \to \infty} e^{\Lstar t} \rhoi = 0 \;.
\end{align}
This means that $H(t) \longrightarrow 0$ as $t \longrightarrow \infty$, which in turn implies that the first term in \eqref{LimtH(t)} has an indeterminate form of the type $\infty \times 0$. We can therefore use L'H$\hat{\rm o}$pital's rule to calculate the limit, giving
\begin{align}
	\lim_{t \to \infty} t \, H(t) = {}& \lim_{t \to \infty} \frac{H(t)}{t^{-1}}  \\
                                = {}& - \lim_{t \to \infty} \frac{\dot{H}(t)}{t^{-2}}  \\
                                = {}& - \lim_{t \to \infty} t^2 \, h(t) = 0  \;,
\end{align}
where the third equality is obtained from \eqref{H(t)}. For the final equality we assume that $h(t)$ is decaying faster than $t^2$ in the limit $t \longrightarrow \infty$, which is the case for most physically relevant probability densities. We note here that it is not always the case that $h(t)$ is such a ``nice'' distribution as this depends on the actual graph defining the quantum walk (e.g.~a graph containing absorbing states may give rise to infinite hitting times). We will not consider such cases in this paper. Equation \eqref{LimtH(t)} thus vanishes completely and the average hitting time becomes
\begin{align}
\label{DefiniteIntegralOfH}
	{\rm E}[\;\!T\;\!] = - \int^\infty_0 \, dt \, H(t) \;.
\end{align}
From \eqref{H(L)} we have
\begin{align}
	F(t) \equiv {}& \int dt \, H(t) \\
               = {}& \sum_{m=1}^{N-1} \, k_{Nm} \, \bra{\psi_m} \, \Lstar^{-1} \left( \int dt \, e^{\Lstar t} \right) \rhoi \ket{\psi_m}   \\
               = {}& \sum_{m=1}^{N-1} \, k_{Nm} \, \bra{\psi_m} \, \big( \Lstar^{-1} \big)^2  \, e^{\Lstar t} \rhoi \;\! \ket{\psi_m}  \;.
\end{align}
The definite integral in \eqref{DefiniteIntegralOfH} is thus given by 
\begin{align}
	{\rm E}[\;\!T\;\!] = - \left[ \lim_{t \to \infty} F(t) - \lim_{t \to 0} F(t) \right]
\end{align}
Using again the fact that $\Lstar$ produces a zero steady state we have 
\begin{align}
	\lim_{t \to \infty} F(t) = 0  \;.
\end{align}
This gives ${\rm E}[\;\!T(\hat{Q}_N,\rhoi)\;\!] = \lim_{t \to 0} F(t)$ which evaluates to
\begin{align}
\label{E[t](L)}
	{\rm E}[\;\!T(\hat{Q}_N,\rhoi)\;\!] = \sum_{m=1}^{N-1} \, k_{Nm} \, \bra{\psi_m} \, \Lstar^{-2}  \, \rhoi \;\! \ket{\psi_m}  \;.
\end{align}
Replacing $\Lstar^{-1}$ by the Moore--Penrose pseudoinverse $\Lstar^+$ then reproduces \eqref{E[t]MPInv}.

\section{Independent derivation of the variance of the hitting time}
\label{AppB}

The variance of the hitting time reads
\begin{align}
	{\rm V}\big[T(\hat{Q}_N,\rhoi)\big] \equiv {\rm E}\big[ T^2(\hat{Q}_N,\rhoi) \big] - \big( {\rm E}\big[ T(\hat{Q}_N,\rhoi) \big] \big)^2  \;.
\end{align}
All we have to do is to calculate ${\rm E}\big[T^2(\rhof,\rhoi)\big]$ which is given by 
\begin{align}
\label{E[T2]}
	{\rm E}\big[T^2(\hat{Q}_N,\rhoi)\big] = \int_0^\infty dt \; t^2 \, h(t;\rhof,\rhoi)  \;.
\end{align}
We can proceed in a similar manner as we have done with the average. Integrating \eqref{E[T2]} by parts we get
\begin{align}
\label{VarStep1}
	{\rm E}\big[T^2(\hat{Q}_N,\rhoi)\big] = \Big[ \, t^2 H(t) \Big|^\infty_0 - 2 \int_0^\infty dt \; t \, H(t)  \;,
\end{align}
where we defined $H(t)$ in \eqref{H(L)}.  Repeated application of the L'H$\hat{\rm o}$pital rule gives
\begin{align}
	 \lim_{t \to \infty} \, t^2 H(t) = - \lim_{t \to \infty} \, t^3  h(t) = 0  \;.
\end{align}
We have assumed again that $h(t)$ is a normalisable function. Using again the fact that $H(t)$ is a continuous function and finite at $t=0$ we get
\begin{align}
	 \lim_{t \to 0} t^2 \, H(t) = 0  \;.
\end{align}
This means that we are left with only the integral in \eqref{VarStep1}.
\begin{align}
\label{VarStep2}
	{\rm E}\big[T^2(\hat{Q}_N,\rhoi)\big] = - 2 \int_0^\infty dt \; t \, H(t) = - 2 \, \bigg\{ \, \Big[ \,t \, F(t)\Big|^\infty_0 - \int_0^\infty dt \, F(t) \bigg\}  \;.
\end{align}
where we have introduced
\begin{align}
	F(t) = \int dt \, H(t)
\label{F(L)}              
       = \sum_{m=1}^{N-1} \, k_{Nm} \, \bra{\psi_m} \, \Lstar^{-2} \, e^{\Lstar t} \rhoi \;\! \ket{\psi_m}  \;.
\end{align}
The finiteness of $F(0)$ and L'H$\hat{\rm o}$pital's rule give once again
\begin{align}
	\Big[ \,t \, F(t)\Big|^\infty_0 = - \lim_{t \to \infty} \, t^2 H(t) -  F(0) \lim_{t \to 0} t \, = 0  \;.
\end{align}
We thus have
\begin{align}
	{\rm E}\big[T^2(\hat{Q}_N,\rhoi)\big] 
	= {}& 2 \int_0^\infty dt \, F(t)  \\
  = {}& 2 \sum_{m=1}^{N-1} \, k_{Nm} \, \bra{\psi_m} \, \Lstar^{-2} \, \int_0^\infty dt \: e^{\Lstar t} \rhoi \;\! \ket{\psi_m}  \\
\label{E[T2]Lbar}                                
  = {}& 2 \sum_{m=1}^{N-1} \, k_{Nm} \, \bra{\psi_m} \, \Lstar^{-3} \, \Big[ e^{\Lstar t} \rhoi \Big|_0^\infty  \;\! \ket{\psi_m}  
	= - 2 \sum_{m=1}^{N-1} \, k_{Nm} \, \bra{\psi_m} \, \Lstar^{-3} \rhoi \ket{\psi_m}  \;,
\end{align}
where we have again used the fact that the dynamics under $\Lstar$ vanishes in the long-time limit. Subtracting off the square of ${\rm E}[T(\hat{Q}_N,\rhoi)]$ we finally arrive at
\begin{align}
	{\rm V}\big[T(\hat{Q}_N,\rhoi)\big] = - 2 \sum_{m=1}^{N-1} \, k_{Nm} \, \bra{\psi_m} \, \Lstar^{-3} \rhoi \ket{\psi_m} 
	                                      -  \Bigg( \sum_{m=1}^{N-1} \, k_{Nm} \, \bra{\psi_m} \, \Lstar^{-2}  \, \rhoi \;\! \ket{\psi_m} \Bigg)^2 \;. 
\end{align}
As with the average, we can replace $\Lstar^{-1}$ by $\Lstar^+$ to obtain \eqref{V[T]Final}.

\end{document}